\pdfoutput=1
\documentclass[aps,prl,twocolumn,superscriptaddress,nofootinbib,
floatfix]{revtex4-2}
\usepackage{amsmath,amssymb}
\usepackage{graphicx}
\usepackage{placeins}
\usepackage[hidelinks]{hyperref}
\newcommand{\Tr}{\mathrm{Tr}}
\newcommand{\E}{\mathbb{E}}

\begin{document}

\author{Jiaxin Liu}
\affiliation{School of Instrumentation and Optoelectronic Engineering,
Beihang University, Haidian, Beijing 100191, China}
\affiliation{Institute of Large-Scale Scientific Facility, Beihang University,
Beijing 100191, China}
\affiliation{Hangzhou Institute of Extremely-Weak Magnetic Field Major
National Science and Technology Infrastructure, Hangzhou 310051, China}
\author{Zuoxian Wang}
\affiliation{School of Instrumentation and Optoelectronic Engineering,
Beihang University, Haidian, Beijing 100191, China}
\affiliation{Institute of Large-Scale Scientific Facility, Beihang University,
Beijing 100191, China}
\author{Feng Li}
\affiliation{Institute of Large-Scale Scientific Facility, Beihang University,
Beijing 100191, China}
\affiliation{Hangzhou Innovation Institute, Beihang University,
Hangzhou 310051, China}
\author{Danyue Ma}
\email{madanyue0419@buaa.edu.cn}
\affiliation{School of Instrumentation and Optoelectronic Engineering,
Beihang University, Haidian, Beijing 100191, China}
\affiliation{Institute of Large-Scale Scientific Facility, Beihang University,
Beijing 100191, China}
\affiliation{Hangzhou Institute of Extremely-Weak Magnetic Field Major
National Science and Technology Infrastructure, Hangzhou 310051, China}
\date{\today}


\title{Record Loss Sets a Rare-Trajectory Limit on Quantum Purification}

\begin{abstract}
Continuous quantum feedback uses time-resolved measurement records to steer
monitored systems toward pure states.  Yet how the information available to a
controller determines the ultimate purification speed remains unresolved.  We
establish this relation for a qubit under fixed-spectrum Hermitian monitoring
with detector loss, obtaining the exact long-time impurity-moment spectrum
optimized over causal basis controls at each horizon.  Rare records with nearly
canceled evidence then make all moments from half order upward decay at the
Bhattacharyya information rate between two quantum nondemolition record laws.
Aligned quantum nondemolition monitoring preserves that binary
distinguishability and attains the limit, while complete detection restores an
order-dependent branch.  The mechanism extends to higher dimensions, where an
attainable rank-two ceiling lies above the full-rank qutrit upper bound over a
finite moment interval, establishing retained record distinguishability as a
purification resource.
\end{abstract}

\maketitle

\def\VTwoFigureOneAnchor{%
\begin{figure}[t!]
  \centering
  \includegraphics[width=\columnwidth]{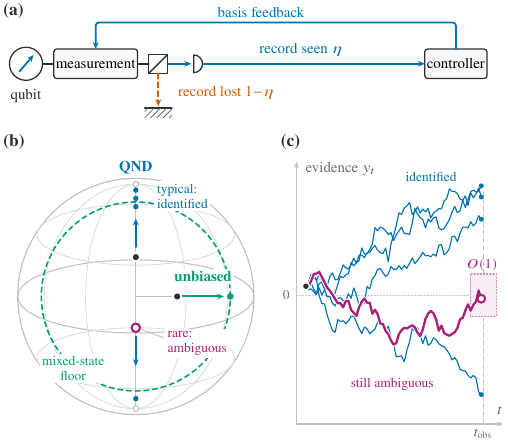}
  \caption{\textbf{Record loss produces a high-order rare-record
  bottleneck.} (a) The full backaction acts on the qubit, while the detected
  fraction \(\eta\) reaches the controller. (b) From the same initial purity,
  unbiased feedback brings the Bloch radius
  \(r_{\rm B}=\sqrt{1-2S}\), with impurity \(S=1-\Tr\rho^2\), to the dashed
  shell \(r_{\rm B}=\sqrt\eta\).
  Typical QND records drive the conditional state toward one pole; the
  highlighted record retains appreciable posterior weight on both
  alternatives. (c) For QND populations
  \(p_\pm\), the binary log-likelihood
  \(y_t=\tfrac12\ln(p_+/p_-)\) grows linearly for typical records.  At
  observation time \(t_{\rm obs}\), the highlighted record has
  \(y_{t_{\rm obs}}=O(1)\) because positive and negative increments nearly
  cancel.}
  \label{fig:mechanism}
\end{figure}%
}


Continuous measurement generates quantum backaction together with a
time-resolved evidence stream, allowing a quantum system to be estimated and
steered during its evolution~\cite{CavesMilburn1987,Belavkin1992,
GisinPercival1992,WisemanMilburn1993,Wiseman1994}.  Quantum-trajectory and
filtering theories
translate that stream into causal state updates and measurement choices,
providing the language for feedback stabilization, rapid readout, error
correction, and autonomous quantum control~\cite{DohertyJacobs1999,GoanEtAl2001,
DohertyJacobsJungman2001,JacobsSteck2006,BoutenVanHandelJames2007,
BarchielliGregoratti2009,WisemanMilburn2010,ZhangEtAl2017,vanHandel2005,
Cardona2020}.  Experiments with superconducting circuits resolve individual
trajectories, verify measurement-induced diffusion, and close feedback loops
in real time on the timescale of the monitored dynamics~\cite{Vijay2012,
Murch2013,Hatridge2013,Campagne2013,Weber2014,
CampagneIbarcq2016,Eddins2019}.  Detection efficiency sets the fraction of
generated evidence available to the controller as the full measurement
interaction continuously shapes the state~\cite{JacobsSteck2006,
WisemanMilburn2010}.  The central question is how usable detected information
sets the attainable purification rate.

Purification speed depends on the ensemble objective, time horizon, and
sequence of measurement orientations chosen by feedback from the evolving
record.  For qubits, unbiased monitoring accelerates average impurity decay,
and quantum
nondemolition (QND) monitoring concentrates probability on rapid first
passage to a prescribed purity threshold~\cite{Jacobs2003,WisemanRalph2006,
WisemanBouten2008}.  Imperfect detection reshapes purification passage times
and finite-time state observables~\cite{Li2013,Jiang2020}, while R\'enyi
terminal costs reveal how the preferred feedback strategy depends on objective
and horizon~\cite{Teo2014}.  The resulting theory spans feedback-accelerated
measurement, higher-dimensional systems, adaptive state preparation,
asymptotic collapse, and multiple information-gain
objectives~\cite{CombesJacobs2006,Griffith2007,CombesWisemanJacobs2008,
ShabaniJacobs2008,BelavkinNegrettiMolmer2009,CWJ2010,
CombesScottWiseman2010,CombesWiseman2011,CombesWiseman2011PRX,Ruskov2012}.
Recent work has sharpened both rate and distributional descriptions of
monitored trajectories.  Bompais \emph{et al.} established exponential
purification in expectation and exponential filter stability for fully
observed finite-dimensional repeated measurements without dark
subspaces~\cite{Bompais2026}.  For~a fixed continuously monitored qubit
initialized at the maximally mixed state, Poltronieri Martins and Lima obtained
an exact path-integral transition law and the resulting state and purity
distributions~\cite{Poltronieri2026}.  Together, these results provide
contraction and propagator benchmarks for monitored trajectories.  Causal basis
control makes the measurement geometry a record-dependent decision variable,
and the long-time impurity-moment spectrum then quantifies how detector loss
constrains that geometry as the statistical emphasis shifts from typical nearly
pure trajectories to rare persistently mixed ones.

In this Letter, we identify rare QND records with nearly canceled evidence as
the source of persistent binary ambiguity under detector loss.  This ambiguity
fixes the high-order branch of the exact qubit spectrum.  A policy-uniform
converse establishes the full spectrum over adaptive orientations, with aligned
QND attaining every moment order under detector loss.  Complete detection
restores the singular rising branch of deterministic unbiased feedback.  The
same binary channel embeds along the extreme monitor eigenstates in any
dimension, carrying the qubit ceiling to rank-two mixtures.  For a full-rank
qutrit, determinant contraction then places its upper bound below that ceiling
over a finite order interval, revealing boundary--interior separation.


\emph{Qubit purification with a lossy record.}
\ifdefined\VTwoPFiveGate\VTwoPFiveGate\fi
Figure~\ref{fig:mechanism}(a) shows that the full interaction acts on the qubit,
while feedback receives the detected fraction of the record.  We set the
channel strength to unity, so rates are quoted per unit dimensionless time.
The conditional qubit state obeys the standard single-channel diffusive
equation~\cite{JacobsSteck2006,WisemanMilburn2010}
\begin{equation}
d\rho_t=\mathcal D[L_t]\rho_t\,dt
 +\sqrt{\eta}\,\mathcal H[L_t]\rho_t\,dW_t ,
\label{eq:sme}
\end{equation}
where \(dW_t^2=dt\), \(0<\eta\leq1\) is the detection efficiency, and
\(L_t=U_tL_0U_t^\dagger\) has fixed measurement contrast
\(2\ell=\lambda_{\max}(L_0)-\lambda_{\min}(L_0)\).  At each time, \(U_t\) is
selected predictably from the detected record available before the next
infinitesimal update.

Inefficiency has a direct beam-splitter interpretation.  After the
qubit--field interaction, one output produces the detected current and the
other carries away an unobserved share of the measurement signal.  The qubit
retains the full backaction, and the controller uses the
\(\sqrt\eta\)-scaled innovation to rotate the subsequent measurement basis.
The lost output enlarges the set of state histories consistent with the
accessible record.  The detected current supplies the likelihood increment
used to update the conditional state and select the next orientation.  Since
the spectrum of \(L_t\) is fixed, its contrast bounds the distinguishability
imprinted per unit dimensionless time; rotation chooses the Bloch direction in
which that distinguishability is spent.  Feedback therefore steers the
geometry of a fixed usable-information budget.  The moment spectrum quantifies
how this geometry distributes purification across common and persistently
ambiguous records.

These record-conditioned basis choices generate a trajectory ensemble whose
still-mixed tail can dominate an ensemble cost.  For impurity
\(S_t=1-\Tr\rho_t^2\), \(S_0>0\), and fixed finite \(\theta>0\), let
\(\Pi_t^{(2)}\) contain all admissible predictable qubit basis rules on
\([0,t]\), including rules designed with the time horizon known, and define
\begin{equation}
\begin{aligned}
V_\theta^{(2)}(t,\eta)
 &=\inf_{\pi\in\Pi_t^{(2)}}\E_\pi[S_t^\theta],\\
\Lambda_\star^{(2)}(\theta,\eta)
 &=\lim_{t\to\infty}-\frac{1}{t}
   \ln V_\theta^{(2)}(t,\eta).
\end{aligned}
\label{eq:objective}
\end{equation}
\ifdefined\VTwoFigureOneAnchor\VTwoFigureOneAnchor\fi
The superscript denotes Hilbert-space dimension, and \(\mathbb E_\pi\)
averages detected records.  Optimization precedes the long-time limit, so the
rule may change with the horizon while remaining causal on its interval.
At fixed time, repeated runs form an impurity distribution.  Its peak follows
the purifying bulk, while its upper tail retains competing state assignments.
Two controls with similar typical trajectories can
therefore have different ensemble rates.  Changing \(S_t^\theta\) shifts the
statistical emphasis across this common distribution.  Since
 \(0\leq S_t\leq1/2\), increasing \(\theta\) reduces the relative contribution
of nearly pure trajectories and emphasizes the progressively rarer mixed tail.
A larger decay exponent corresponds to faster asymptotic purification.  The
dependence of \(\Lambda_\star^{(2)}\) on \(\theta\) provides a moment-resolved
diagnostic; a plateau signals that one rare sector controls many moment orders.

The endpoints in Fig.~\ref{fig:mechanism}(b) allocate the record differently.
Unbiased monitoring makes the Bloch radius deterministic and leaves a
loss-dependent mixed-state floor~\cite{Jacobs2003,CombesWiseman2011}.  QND
monitoring repeatedly probes the same observable.  The two long-time laws are
\begin{equation}
\begin{aligned}
S_t^{\rm ub}&\longrightarrow \frac{1-\eta}{2}
 &&(0<\eta<1),\\
\Pr_{\rm QND}(S_t\ge s)
 &=e^{-2\eta\ell^2t+o(t)}
 &&\left(0<s<\tfrac12,\ 0<\eta\leq1\right).
\end{aligned}
\label{eq:endpoint-contrast}
\end{equation}
\begin{samepage}
The first law describes convergence of the trajectory bulk to its
loss-dependent shell.  The
second gives the probability of remaining appreciably mixed under QND\@.  These
strategies allocate the same detected information differently: the unbiased
basis contracts the bulk toward the mixed-state shell, while QND transfers
probability toward the poles and retains an exponentially sparse unresolved
sector.  Their competition is therefore decided by the ensemble objective.
For any mixed aligned initial state, the fixed-QND propagator fixes the
threshold law and its prefactor~\cite{Li2013,Jiang2020,SM}.  Its maximally mixed
specialization agrees with the exact path-integral transition law of
Ref.~\cite{Poltronieri2026}.  Optimizing the moment spectrum determines which
allocation sets each order over all causal basis choices.
\end{samepage}

For every mixed initial qubit, the optimized limit in
Eq.~\eqref{eq:objective} exists and equals
\begin{equation}
\Lambda_\star^{(2)}(\theta,\eta)=
\begin{cases}
8\theta(1-\theta)\eta\ell^2,
 &0<\theta<\tfrac12,\quad 0<\eta\le1,\\[2pt]
2\eta\ell^2,
 &\theta\ge\tfrac12,\quad 0<\eta<1,\\[2pt]
4\theta\ell^2,
 &\theta\ge\tfrac12,\quad \eta=1.
\end{cases}
\label{eq:qubit-spectrum}
\end{equation}
Equation~\eqref{eq:qubit-spectrum} is the main result of this Letter.  It gives
the long-time exponent of the finite-horizon optimum itself, allowing the
causal basis policy to be redesigned at every time horizon.  At partial
efficiency, QND realizes the spectrum, and all \(\theta\geq1/2\) share the
rare-record cost \(2\eta\ell^2\).  Proving QND optimality therefore requires
showing that arbitrary adaptive rotations retain this unresolved sector.
The same rare-record transition governs long-time mean R\'enyi entropies.
Near purity, entropy of order \(\alpha\) has the exponential scale
\(S^{q_\alpha}\), where \(q_\alpha=\min\{\alpha,1\}\).
Equation~\eqref{eq:qubit-spectrum} therefore places every
\(\alpha\geq1/2\) on the lossy plateau, while the ideal detector retains the
rising branch~\cite{SM}.

\ifdefined\VTwoStopAfterHero
  \expandafter\endinput
\fi

\emph{Mechanism of spectral freezing.}
The plateau originates in large deviations of the QND log-likelihood.  Write
the two QND populations as \(p_\pm\) and define the
binary log-likelihood coordinate \(2y=\ln(p_+/p_-)\), for which
\(S=\tfrac12\operatorname{sech}^2y\).  Resolve the detected record into short
time bins, each slightly favoring one eigenstate.  A typical sequence develops
a linear signed imbalance, so \(|y_t|\) grows steadily and the conditional
state approaches a pole.  In a rare sequence, positive and negative increments
remain nearly balanced.  Its terminal \(y_t\) stays \(O(1)\), and the
conditional state retains two plausible alternatives after a long observation
[Fig.~\ref{fig:mechanism}(c)].  The moment average balances the impurity
associated with a chosen log-likelihood velocity against the probability cost
of producing that velocity.  The probability of such balanced histories decays
exponentially while their conditional impurity remains appreciable.  Below
\(\theta=\tfrac12\), the selected velocity
is nonzero and the conditional states purify exponentially.  It reaches zero
at the half moment.  Every higher order selects bounded-\(y\) records with the
same exponential cost \(2\eta\ell^2\)~\cite{SM}.  Spectral freezing is the rate
signature of this large-deviation selection.

The half moment gives this rate an exact record-law meaning.  Let
\(\mathsf P_\pm^{(t)}\) be the detected-record path laws conditioned on the two
monitor eigenstates.  Bayes' rule identifies the normalized half moment with
their Bhattacharyya coefficient, equivalently their Hellinger affinity.  For
every mixed aligned initial qubit,
\begin{equation}
\begin{aligned}
\frac{\E_{\rm QND}[\sqrt{S_t}]}{\sqrt{S_0}}
 &=\mathcal B_t
 :=\int\sqrt{d\mathsf P_+^{(t)}d\mathsf P_-^{(t)}}\\
 &=e^{-2\eta\ell^2t}.
\end{aligned}
\label{eq:pivot-identity}
\end{equation}
Thus \(-t^{-1}\ln\mathcal B_t=2\eta\ell^2\) is the detected-record
Bhattacharyya information rate of the QND record pair.  This affinity is the
continuous-time QND counterpart of the qubit contraction coefficient appearing
in general repeated-measurement purification bounds~\cite{Bompais2026}.  The
equality of this information rate with the frozen plateau identifies the
discrimination rate relevant to persistent binary ambiguity.  The half moment
is therefore the junction between the moving-velocity saddle and the
bounded-\(y\) sector.

\def\VTwoFigTwoAnchor{%
\begin{figure}[t!]
  \centering
  \includegraphics[width=\columnwidth]{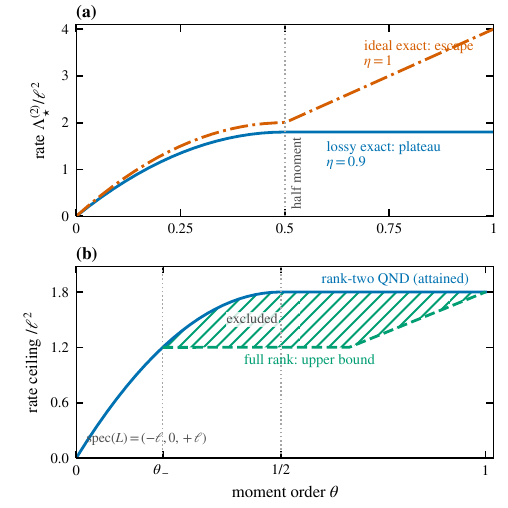}
  \caption{\textbf{Exact qubit spectrum and full-rank qutrit bound.}
  (a) Qubit optimum at \(\eta=0.9\) (blue) and \(\eta=1\) (orange).
  (b) At \(\eta=0.9\) for
  \(L=\operatorname{diag}(-\ell,0,\ell)\), rank-two QND attains the
  qubit ceiling; the dashed full-rank qutrit upper bound lies strictly below
  this attainable ceiling for \(\theta_-<\theta<1\), with
  \(\theta_-=(1-1/\sqrt3)/2\).}
  \label{fig:spectrum}
\end{figure}%
}


Adaptive orientations redistribute the fixed spectral contrast while
preserving its width.  Two rotation-invariant budgets yield a policy-uniform
converse and meet at the half moment, linking the trace-inequality control of
low orders to the log-impurity control of slowly purifying high-order
trajectories.

For \(\theta\leq1/2\), rotating the measurement axis distributes the fixed
contrast between deterministic contraction and distinguishability in the
detected record.  Quantitatively, for every admissible causal qubit basis
policy driven by that record, \(0<\eta\leq1\),
\(0<\theta\leq\tfrac12\), and \(t\geq0\),
\begin{equation}
\E_\pi\bigl[S_t^\theta\bigr]\geq
S_0^\theta\,e^{-8\theta(1-\theta)\eta\ell^2 t}.
\label{eq:low-order-converse}
\end{equation}
Together with QND attainment, Eq.~\eqref{eq:low-order-converse} fixes the
curved branch in Eq.~\eqref{eq:qubit-spectrum}.  At \(\theta=1/2\),
Eq.~\eqref{eq:pivot-identity} saturates the bound at every time and anchors the
transition to bounded-\(y\) QND records.  The equality geometry of the sharp
trace inequality selects extremal QND subspaces in higher
dimensions~\cite{SM}.

For \(0<\eta<1\) above the half moment, log impurity isolates a near-pure
corridor of slowly purifying trajectories.  Fixed spectral width bounds both its
residual radial drift and its quadratic variation.  A horizon-indexed change
of measure assigns order-one probability to this corridor, while the
likelihood-ratio cost under the physical record law is bounded at rate
\(2\eta\ell^2\), uniformly over policies and time horizons.  The optimized
moment therefore has the same high-order ceiling, attained by bounded-\(y\)
QND records.  At complete efficiency, deterministic unbiased flow saturates
the finite-time envelope.  The stopped-event certificate retains uniform
constants through the long-time and vanishing-width limits~\cite{SM}.

\ifdefined\VTwoPThirteenGate\VTwoPThirteenGate\fi
The ideal detector is a singular endpoint of the policy-uniform ceiling.  For
every \(0<\eta<1\), rare inconclusive QND
records set the optimized high-order rate \(2\eta\ell^2\), while unbiased
feedback settles at the floor \((1-\eta)/2\).  As \(\eta\uparrow1\), the floor
vanishes and the crossing time \(t_\times\), defined by
\(\E_{\rm QND}[S_{t_\times}]=S_{t_\times}^{\rm ub}\), grows logarithmically for
fixed \(S_0>0\).  This creates a widening preasymptotic window of
complete-detection behavior.  At \(\eta=1\), the complete record enables
unbiased feedback to convert each update into deterministic radial contraction
at rate \(4\theta\ell^2\).  Thus, for each fixed
\(\theta>\tfrac12\),
\(\lim_{\eta\to1^-}\Lambda_\star^{(2)}(\theta,\eta)=2\ell^2\), and
\(\Lambda_\star^{(2)}(\theta,1)=4\theta\ell^2\)
[Fig.~\ref{fig:spectrum}(a)].
\ifdefined\VTwoPThirteenEndGate\VTwoPThirteenEndGate\fi


\clearpage
\ifdefined\VTwoFigTwoAnchor\VTwoFigTwoAnchor\fi
\emph{Higher dimensions.}
The qubit mechanism embeds into every larger Hilbert space.  The
eigenvectors of \(L\) with largest and smallest eigenvalues span a
two-dimensional subspace that uses the full monitor contrast.  A commuting
mixture supported on this extremal subspace remains there under QND monitoring:
the resulting binary channel repeatedly compares the same two alternatives,
thereby reproducing the qubit spectrum and attaining its dimension-independent
ceiling for \(0<\eta<1\).  The same scalar log-impurity corridor yields the
policy-uniform bound in the ambient state space.
Let \(V_\theta^{(d)}\) be the \(d\)-dimensional analogue of
Eq.~\eqref{eq:objective}, and let \(\Lambda_\star^{(d)}\) denote its long-time
liminf rate.  Fix finite \(\theta>0\) and \(0<\eta<1\).  Every mixed initial
state in finite \(d\geq2\) obeys the dimension-independent ceiling, which the
extremal-pair embedding attains.  For a full-rank qutrit, \(\det\rho_0>0\),
monitored by \(L=\operatorname{diag}(-\ell,0,\ell)\), determinant contraction
gives the additional upper bound
\ifdefined\VTwoPFourteenGate\VTwoPFourteenGate\fi
\begin{equation}
\begin{aligned}
\Lambda_\star^{(d)}(\theta,\eta)
  &\leq \Lambda_\star^{(2)}(\theta,\eta),\\
\Lambda_\star^{(3)}(\theta,\eta)
  &\leq \frac{4}{3}\eta\ell^2
  \max\!\left\{1,\frac{3\theta}{2}\right\}.
\end{aligned}
\label{eq:higher-dimensional-consequence}
\end{equation}
\ifdefined\VTwoEFiveGate\VTwoEFiveGate\fi
Extremal rank-two QND mixtures attain the dimension-independent ceiling.  The
full-rank qutrit upper bound is strictly lower for
\((1-1/\sqrt3)/2<\theta<1\).

The extremal subspace concentrates the full contrast into binary evidence.
For a qutrit, \(R=(\det\rho)^{1/3}\) is the geometric mean of its three
eigenvalues; \(R>0\) exactly for a full-rank state in the interior of state
space and measures genuine three-level spectral volume.  The determinant law
supplies this stricter upper bound, so the boundary mixture and the full-rank
state probe two geometries generated by one usable-information budget.  The
qubit spectrum thereby becomes a benchmark for higher-dimensional
trajectories.

The same rare-record mechanism determines how the predicted spectrum can be
tested.  Under aligned QND, repeated calibrated records provide filtered state
estimates at several observation times.  For fixed \(\theta>1/2\), the
spectral-edge asymptotics give
\(\operatorname{Var}(S_t^\theta)/\E[S_t^\theta]^2
\asymp t^{1/2}e^{2\eta\ell^2t}\).  Direct independent sampling therefore
requires
\(N_{\rm traj}\asymp\epsilon^{-2}t^{1/2}e^{2\eta\ell^2t}\) for relative
root-mean-square error \(\epsilon\).  A tilted record ensemble can make the
bounded-evidence histories typical, with likelihood-ratio reweighting restoring
the physical moment.  The same likelihood tilt supplies a natural
importance-sampling proposal~\cite{SM}.
These sampling requirements become most demanding near complete detection,
where the observation window must extend beyond the logarithmically receding
crossing between the QND and unbiased mean-impurity curves.

\emph{Discussion and conclusion.}
Under fixed-spectrum monitoring with detector loss, the attainable
impurity-moment spectrum is set by the distinguishability retained in the
detected record.  Causal basis rotations redistribute the fixed spectral
contrast among trajectories, while evidence lost at the detector remains
inaccessible.  That asymmetry leaves a rare sector of nearly canceled records,
whose Bhattacharyya information rate fixes the optimized high-order decay.
Aligned QND monitoring preserves the binary comparison and attains that rate,
whereas complete detection makes the full evidence available to deterministic
unbiased feedback.  For partial efficiency, the same lossy binary channel
embeds in every dimension through the extremal monitor eigenstates.  Its
rank-two ceiling is attainable, while determinant contraction places the
full-rank qutrit upper bound strictly below it over a finite moment interval.
Operationally, eigenstate calibrations determine the record-law overlap, while
mixed-state runs estimate long-time qubit and qutrit slopes against these
bounds.  Tilted rare-event estimates complement these calibrated measurements
and test the predicted link between record distinguishability and purification
speed.
\ifdefined\VTwoPSeventeenEndGate\VTwoPSeventeenEndGate\fi

\begin{acknowledgments}
\emph{Acknowledgments.}
We thank Howard M. Wiseman for valuable discussions and detailed comments that
substantially improved the physical presentation of this work. This work was
supported by the National Natural Science Foundation of China under Grant
No.~42388101.
\end{acknowledgments}

\emph{Data availability.} The numerical data and custom code supporting the
figures are available from the authors upon reasonable request.




\clearpage
\onecolumngrid
\section*{End Matter}
\label{sec:v2-endmatter}
\begin{figure}[!t]
\centering
\includegraphics[width=0.985\textwidth]{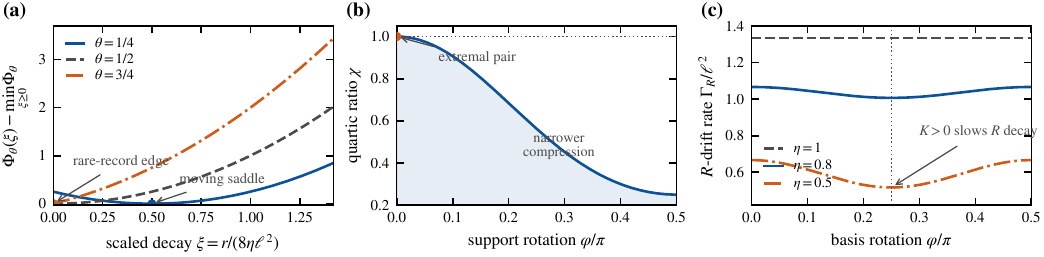}
\caption{\textbf{Proof geometry behind the speed limit.}
(a) Under aligned QND, with \(\xi=r/(8\eta\ell^2)\),
\(\Phi_\theta=(\xi-1)^2+4\theta\xi\) is minimized at
\(\xi_\theta=\max\{1-2\theta,0\}\), reaching the rare-record edge at
\(\theta=1/2\).  (b) For \(L=\ell\operatorname{diag}(-1,0,1)\), a rotated
rank-two support at angle \(\varphi\) has
\(\chi=(T_2S+T_3^2)/(\ell^2S^2)<1\) away from the extremal pair, where
\(T_2=\Tr(\widetilde L\rho\widetilde L\rho)\) and
\(T_3=\Tr(\rho^2\widetilde L)\).  (c) For the same \(L\) and
  \(\rho=\operatorname{diag}_{(-,0,+)}(0.50,0.30,0.20)\), an
  extremal-subspace rotation by \(\varphi\) gives
  \(\Gamma_R=2\eta V_L-(1-\eta)K/3\), with
  \(V_L=\Tr[(L-\tfrac13\Tr L\,\openone)^2]/3\) and
  \(K=\Tr(\rho^{-1}L\rho L)-\Tr L^2\).  For \(\eta<1\), noncommuting
  rotations have \(K>0\) and slow the decay of \(R=(\det\rho)^{1/3}\).  At
  \(\eta=1\), the rate is orientation independent.}
\label{fig:endmatter-proof}
\end{figure}

\noindent\begin{minipage}[t]{0.485\textwidth}
\vspace{0pt}

\emph{Model and control class.}
Throughout the Letter, \(t\) denotes dimensionless time with the channel
strength set to unity, and \(dW_t^2=dt\).  This convention fixes resource
units.  In Eq.~\eqref{eq:sme},
\(L_t=U_tL_0U_t^\dagger\) is Hermitian, and every \(L_t\) has the fixed
measurement contrast
\(2\ell=\lambda_{\max}-\lambda_{\min}\).  For a generic instantaneous
orientation \(L\),
\begin{equation}
\begin{aligned}
 \mathcal D[L]\rho&=L\rho L-\tfrac12\{L^2,\rho\},\\
 \mathcal H[L]\rho&=L\rho+\rho L-2\Tr(\rho L)\rho.
\end{aligned}
\label{eq:methods-model}
\end{equation}
Both superoperators are invariant under scalar shifts of \(L\).  We consider
one finite-dimensional Hermitian diffusive channel with fixed spectrum,
\(\ell>0\), \(0<\eta\le1\), and mixed \(S_0>0\).  The controller chooses
\(U_t\) predictably from the detected past, with unrestricted basis bandwidth.
A predictable, locally time-integrable Hamiltonian is absorbed into this
relative orientation because the finite-variation evolution generated by a
Hamiltonian preserves the state spectrum.  For each time horizon \(t\), the
class \(\Pi_t^{(d)}\) contains all such predictable policies on \([0,t]\),
including policies designed with that horizon known.  Rates are per unit
dimensionless time; the Supplemental Material (SM) restores laboratory time.

\emph{Qubit reduction.}
In the instantaneous eigenbasis of a qubit state, put
\(S=2\det\rho\), \(\beta=1-2S\), and
\(x=(L_{11}-L_{22})^2/4\in[0,\ell^2]\).  Direct It\^o evaluation gives
\begin{equation}
\begin{aligned}
 dS={}&-2g(x,S)\,dt
       -4S\sqrt{\eta x\beta}\,dW_t,\\
 g(x,S)={}&(\eta-1)
  \big[(\ell^2-x)(1-S)+2xS^2\big]\\
 &+(1+\eta)S\big(\ell^2-x\beta\big).
\end{aligned}
\label{eq:methods-qubit}
\end{equation}
The QND and unbiased orientations are \(x=\ell^2\) and \(x=0\), respectively.
Both belong to the control class; the latter can be selected predictably from
the pre-innovation state.  For any twice-differentiable scalar cost \(F(S)\),
its generator is affine in \(x\).  Endpoint inequalities therefore extend to
every intermediate orientation and reduce each local moment comparison to the
two extremes.

\emph{Exactly solvable endpoints.}
The unbiased endpoint has deterministic radial flow,
\begin{equation}
 S_t^{\rm ub}=\frac{1-\eta}{2}
 +\left(S_0-\frac{1-\eta}{2}\right)e^{-4\ell^2t}.
\label{eq:methods-unbiased}
\end{equation}
The aligned QND endpoint preserves diagonality.  With populations \(p_\pm\),
its binary log-likelihood \(2y=\ln(p_+/p_-)\) satisfies
\(S_t=\tfrac12\operatorname{sech}^2y_t\) and obeys the diffusion
\begin{equation}
 dy_t=4\eta\ell^2\tanh y_t\,dt
       +2\ell\sqrt\eta\,dW_t.
\label{eq:methods-endpoint-dynamics}
\end{equation}
Its half moment is the exact identity in Eq.~\eqref{eq:pivot-identity}, and
the aligned threshold law in Eq.~\eqref{eq:endpoint-contrast} follows from
the same kernel.

\end{minipage}\hspace{0.03\textwidth}%
\begin{minipage}[t]{0.485\textwidth}
\vspace{0pt}

\emph{QND moment spectrum.}
For the aligned QND endpoint, the exact long-time moment exponent is
\begin{equation}
 \Lambda_{\rm QND}(\theta,\eta)
 =\begin{cases}
 8\theta(1-\theta)\eta\ell^2,&0<\theta\le\tfrac12,\\
 2\eta\ell^2,&\theta\ge\tfrac12
 \end{cases}.
\label{eq:methods-endpoints}
\end{equation}
Conjugating the QND generator with \(\cosh y\) leaves free
heat flow with spectral shift \(2\eta\ell^2\) and transformed weight
\(\cosh^{1-2\theta}y\).  Below \(\theta=1/2\), a moving Gaussian saddle
dominates.  At the half moment it reaches the zero-decay rare-record edge and
the weight becomes constant, yielding Eq.~\eqref{eq:pivot-identity}
[Fig.~\ref{fig:endmatter-proof}(a)].  The figure uses the realized decay rate
\(r=-t^{-1}\ln S_t\).  Near-pure entropy--impurity comparison transfers the
spectrum to mean R\'enyi entropy through
\(q_\alpha=\min\{\alpha,1\}\).

\emph{Finite-time envelope.}
For arbitrary finite dimension, center the observable,
\(\widetilde L=L-\Tr(\rho L)\openone\), and collect all orientation
dependence into the three scalars \(T_1=\Tr(\widetilde L^2\rho^2)\),
\(T_2=\Tr(\widetilde L\rho\widetilde L\rho)\), and
\(T_3=\Tr(\rho^2\widetilde L)\).  The scalar \(T_2\) enters the radial drift,
whereas \(T_3\) controls the radial innovation.  The difference
\(T_1-T_2=\tfrac12\|[\widetilde L,\rho]\|_F^2\ge0\) measures
noncommutativity.  It\^o's formula gives
\begin{equation}
 dS=-2\big[(1+\eta)T_2-(1-\eta)T_1\big]dt
 -4\sqrt\eta\,T_3\,dW_t.
\label{eq:methods-impurity}
\end{equation}
The two sharp spectral inequalities
\begin{equation}
 T_2S+T_3^2\le\ell^2S^2,\qquad |T_3|\le\ell S,
\label{eq:methods-spectral}
\end{equation}
combine into the master inequality
\begin{equation}
 (1+\eta)T_2-(1-\eta)T_1
 +2\eta\frac{T_3^2}{S}\le2\eta\ell^2S.
\label{eq:methods-master}
\end{equation}
Its left-hand side equals
\(2\eta(T_2+T_3^2/S)-(1-\eta)(T_1-T_2)\), displaying the competition between
radial contraction and noncommutativity.  Writing \(\mathcal A\) for the
controlled generator, Eq.~\eqref{eq:methods-master} reduces every impurity
moment to
\begin{equation}
 -\frac{\mathcal A S^\theta}{S^\theta}
 \le4\theta\eta\ell^2
 +4\theta(1-2\theta)\eta\frac{T_3^2}{S^2}.
\label{eq:methods-generator}
\end{equation}
For \(0<\theta\le1/2\), the bound \(|T_3|\le\ell S\) gives the first branch; for
\(\theta\ge1/2\), the residual term is nonpositive.  An exponential transform
and a standard stopping-time argument then yield, for every \(t\ge0\) and every
admissible predictable policy,
\begin{equation}
\begin{aligned}
 \E_\pi[S_t^\theta]&\ge S_0^\theta
   e^{-c(\theta)\eta\ell^2t},\\
 c(\theta)&=\begin{cases}
 8\theta(1-\theta),&0<\theta\le\tfrac12,\\
 4\theta,&\theta\ge\tfrac12,
 \end{cases}
\end{aligned}
\label{eq:methods-envelope}
\end{equation}
whose low-order case is Eq.~\eqref{eq:low-order-converse} and whose
\(\theta=1/2\) case is saturated by Eq.~\eqref{eq:pivot-identity}.
\end{minipage}\hfill\par

\clearpage
\setlength{\textfloatsep}{5pt plus 2pt minus 1pt}
\setlength{\abovecaptionskip}{4pt}

\noindent\begin{minipage}[t]{0.485\textwidth}
\vspace{0pt}

\emph{Equality and dimensional embedding.}
For \(S>0\), the first inequality in Eq.~\eqref{eq:methods-spectral} is
saturated exactly when \(\rho\) has rank two and the compression of \(L\) to
\(\operatorname{supp}\rho\) has width \(2\ell\)
[Fig.~\ref{fig:endmatter-proof}(b)].  At \(0<\eta<1\), saturation of the
half-moment bound additionally requires QND commutation, except at the
maximally mixed qubit.  An extremal commuting mixture thus embeds the low-order
QND exponent in any \(d\).  At \(\eta=1\) and \(d=2\), \(S=2\det\rho\) makes
the critical-order bound independent of the predictable basis history,
yielding basis-independent critical-order dynamics.

\emph{The inefficient high-order ceiling.}
For \(\theta>1/2\) and \(0<\eta<1\), the long-time rate sharpens the
finite-time envelope.  Set \(\Delta=T_1-T_2\),
\(B=2\eta T_2-(1-\eta)\Delta\), and
\(s_\eta=\min\{1/4,(1-\eta)/(16\eta)\}\).  The spectral inequalities give
\(B\le24\eta\ell^2S^2\) for \(S\le s_\eta\).  In the near-pure region
\(S\le s<s_\eta\), let \(\chi_u=\mathbf1_{\{S_{u-}\le s\}}\),
\(X=\tfrac12\ln S\), and \(\nu=-2\sqrt\eta\,T_3/S\).  Then
\begin{equation}
 dX=-\left(\frac BS+\nu^2\right)dt+\nu\,dW_t,\qquad
 \nu^2\le4\eta\ell^2.
\label{eq:methods-corridor}
\end{equation}
The quadratic-variation budget obeys
\(\tfrac12\int_0^t\chi_u\nu_u^2\,du\le2\eta\ell^2t\).
The estimate on \(B\) bounds the residual drift \(-B/S\) from below by
\(-24\eta\ell^2s\).  A change of measure cancels the \(-\nu^2\) term.
A stopping-time argument gives a bound whose constants are independent of the
policy in \(\Pi_t^{(d)}\).  Hence the policy infimum may be taken before the
long-time limit, yielding
\begin{equation}
 \limsup_{t\to\infty}-\frac1t
 \ln\!\left(\inf_{\pi\in\Pi_t^{(d)}}\E_\pi[S_t^\theta]\right)
 \le2\eta\ell^2 ,
\label{eq:methods-converse}
\end{equation}
for fixed finite \(d\), mixed \(S_0>0\), and \(\theta>1/2\).  The limit is
taken first at fixed corridor width and then as that width vanishes.  For
qubits, the stopped-event certificate is uniform, and QND attains the ceiling.
\end{minipage}\hspace{0.03\textwidth}%
\begin{minipage}[t]{0.485\textwidth}
\vspace{0pt}

\emph{Full-rank determinant law.}
For \(\det\rho_0>0\), set \(\mu=\Tr L/d\),
\(V_L=\Tr(L-\mu\openone)^2/d\), \(R=(\det\rho)^{1/d}\), and
\(K=\Tr(\rho^{-1}L\rho L)-\Tr L^2\ge0\).  On the full-rank interval,
matrix It\^o calculus gives
\begin{equation}
 \frac{dR}{R}
 =\left[-2\eta V_L+\frac{1-\eta}{d}K(\rho,L)\right]dt
 +2\sqrt\eta\big(\mu-\Tr\rho L\big)dW_t.
\label{eq:methods-detsde}
\end{equation}
In an eigenbasis of \(\rho\), \(K\) is a sum of nonnegative terms.
Noncommutation therefore reduces the conditional decay of \(R\) for
\(\eta<1\) [Fig.~\ref{fig:endmatter-proof}(c)].  Hamiltonian rotations
preserve \(R\).  A standard stopping-time argument shows that
\(e^{2\eta V_Lt}R_t\) is a true submartingale.  Consequently,
\begin{equation}
 \E_\pi R_t\ge R_0e^{-2\eta V_Lt}.
\label{eq:methods-detroot}
\end{equation}
At \(\eta=1\), equality holds for every policy.  At \(0<\eta<1\), equality
holds whenever \([\rho_t,L_t]=0\) for \(dt\otimes d\mathbb P\)-almost every
\((t,\omega)\).

\emph{From determinant to impurity.}
Let \(\gamma_d=(d-1)/d\) and \(c_d=[2(d-1)/d]^{\gamma_d}\).  The eigenvalue
comparison \(S^{\gamma_d}\ge c_dR\), together with monotonicity for
\(\theta\le\gamma_d\) and Jensen's inequality for \(\theta>\gamma_d\), gives
\begin{equation}
 \Lambda_\star^{(d)}(\theta,\eta)
 \le2\eta V_L\max\!\left\{1,\frac{\theta}{\gamma_d}\right\}.
\label{eq:methods-detceiling}
\end{equation}
For \(d=3\) and \(\operatorname{spec}L=(-\ell,0,\ell)\),
\(V_L=2\ell^2/3\), which reproduces
Eq.~\eqref{eq:higher-dimensional-consequence}.  For \(0<\eta<1\), this qutrit
bound lies below the attainable rank-two ceiling
when \((1-1/\sqrt3)/2<\theta<1\) [Fig.~\ref{fig:spectrum}], proving strict
boundary--interior separation throughout this interval.  The same spectral
comparison gives the corresponding interval for a general spectrum~\cite{SM}.
\end{minipage}\hfill\par

\clearpage
\onecolumngrid
\setcounter{secnumdepth}{3}
\setcounter{section}{0}
\setcounter{subsection}{0}
\setcounter{equation}{0}
\setcounter{figure}{0}
\setcounter{table}{0}
\setcounter{footnote}{0}
\renewcommand{\theequation}{S\arabic{equation}}
\renewcommand{\theHequation}{SM.S\arabic{equation}}
\renewcommand{\thesection}{\Roman{section}}
\renewcommand{\theHsection}{SM.\arabic{section}}
\renewcommand{\thesubsection}{\thesection.\Alph{subsection}}
\renewcommand{\theHsubsection}{SM.\arabic{section}.\arabic{subsection}}
\renewcommand{\thetable}{S\Roman{table}}
\renewcommand{\theHtable}{SM.S\arabic{table}}
\renewcommand{\thefigure}{S\arabic{figure}}
\renewcommand{\theHfigure}{SM.S\arabic{figure}}

\begin{center}
{\large\bfseries Supplemental Material for\\[3pt]
Record Loss Sets a Rare-Trajectory Limit on Quantum Purification}

\medskip
{\normalsize Jiaxin Liu, Zuoxian Wang, Feng Li, and Danyue Ma$^{*}$}\\[3pt]
{\small $^{*}$Corresponding author:
\href{mailto:madanyue0419@buaa.edu.cn}{madanyue0419@buaa.edu.cn}}
\end{center}
\medskip

This Supplemental Material follows the proof dependencies of the Letter.  We
first reduce the qubit to a single orientation variable and solve its quantum
nondemolition (QND) channel, making the rare-record mechanism explicit.  We
next transfer the long-time moment spectrum to R\'enyi entropies.  We then
establish the trace inequalities, finite-time moment envelope, inefficient
long-time ceiling, qubit event certificate, and full-rank determinant
separation.  The final sections collect secondary finite-time resolutions.
Throughout Secs.~\ref{sm:param}--\ref{sm:secondary}, \(t\) denotes dimensionless
time with the measurement strength set to unity.
Section~\ref{sm:labunits} restores laboratory-time units.

For each finite dimension \(d\) and time horizon \(t\), let
\(\Pi_t^{(d)}\) denote the admissible predictable fixed-spectrum orientation
rules on \([0,t]\).  A rule may be designed with \(t\) known while remaining
causal with respect to the detected record.  For a fixed mixed initial state,
define
\begin{equation}
\begin{aligned}
 V_\theta^{(d)}(t,\eta)
 &=\inf_{\pi\in\Pi_t^{(d)}}\E_\pi[S_t^\theta],\\
 \Lambda_\star^{(d)}(\theta,\eta)
 &=\liminf_{t\to\infty}-\frac1t
   \ln V_\theta^{(d)}(t,\eta).
\end{aligned}
\label{eq:sm-horizon-value}
\end{equation}

\section{Qubit reduction and endpoint dynamics}
\label{sm:param}

\subsection{Orientation and invariants}

Write the qubit state in its instantaneous eigenbasis as
\(\rho=\operatorname{diag}(\lambda,1-\lambda)\), with
\(\lambda\geq\tfrac12\).  Its impurity and spectral gap are
\[
S=1-\Tr\rho^2=2\lambda(1-\lambda)=2\det\rho,
\qquad \beta=(2\lambda-1)^2=1-2S.
\]
The controlled observable is
\(L_t=U_tL_0U_t^\dagger\), where \(U_t\) is predictable and every \(L_t\)
has the same spectrum.  For a qubit, a scalar shift leaves the dynamics
invariant, so the fixed eigenvalues may be taken as \(\pm\ell\).  In the
instantaneous state eigenbasis we suppress the time subscript and write
\begin{equation}
 |L_{12}|^2+\Delta^2/4=\ell^2,
 \qquad \Delta=L_{11}-L_{22},
 \label{eq:sm-identity}
\end{equation}
and the relative orientation is completely specified by
\(x=\Delta^2/4\in[0,\ell^2]\).  The endpoints \(x=\ell^2\) and \(x=0\)
are respectively the aligned QND and unbiased bases.  With
\(\widetilde L=L-\Tr(\rho L)\openone\), direct evaluation gives
\begin{align}
 \Tr(\widetilde L\rho\widetilde L\rho)
 &=S(\ell^2-x\beta),
 \label{eq:sm-inv1}\\
 \Tr(\widetilde L^2\rho^2)
 &=(\ell^2-x)(1-S)+2xS^2,
 \label{eq:sm-inv2}\\
 [\Tr(\widetilde L\rho^2)]^2
 &=xS^2\beta.
 \label{eq:sm-inv3}
\end{align}
For example, the diagonal and off-diagonal parts of the first trace are
\(2xS^2\) and \(S(\ell^2-x)\), respectively.  Their sum is
\(S[\ell^2-x(1-2S)]\).  The remaining identities follow analogously.

\subsection{Impurity dynamics and the unbiased endpoint}

For \(Q=\Tr\rho^2\), It\^o's formula gives
\(dQ=2\Tr(\rho\,d\rho)+\Tr[(d\rho)^2]\).  Finite-variation Hamiltonians
preserve the state spectrum and are absorbed into the relative orientation.
The measurement terms reduce through
\[
2\Tr(\rho\mathcal D[L]\rho)
=2\!\left[
\Tr(\widetilde L\rho\widetilde L\rho)
-\Tr(\widetilde L^2\rho^2)\right],
\]
\[
\Tr[(\mathcal H[L]\rho)^2]
=2\!\left[
\Tr(\widetilde L^2\rho^2)
+\Tr(\widetilde L\rho\widetilde L\rho)\right],
\]
whose qubit values are given in
Eqs.~\eqref{eq:sm-inv1}--\eqref{eq:sm-inv2}.  Consequently,
\begin{equation}
\begin{aligned}
 dS={}&-2g(x,S)\,dt-4S\sqrt{\eta x\beta}\,dW_t,\\
 g(x,S)={}&(\eta-1)
 \big[(\ell^2-x)(1-S)+2xS^2\big]\\
 &+(1+\eta)S(\ell^2-x\beta).
\end{aligned}
\label{eq:sm-sde}
\end{equation}
Both the drift and squared diffusion are affine in \(x\).  Thus, for every
\(C^2\) scalar cost \(F(S)\), the conditional generator
\(\mathcal AF(S)\) is affine in \(x\), where \(\mathcal A\) denotes the
conditional infinitesimal generator at the selected orientation.  A pointwise
endpoint inequality then holds for any predictable intermediate orientation.
The admissible control is an arbitrary predictable fixed-spectrum orientation
\(L_t\), or equivalently \(x_t\in[0,\ell^2]\).  For these spectral costs, a
separate locally time-integrable Hamiltonian changes the same relative
orientation.  At each time, the controller may select \(L_t\) predictably
from the pre-innovation state to impose \(x_t=0\).  The always-unbiased
endpoint is therefore included within predictable orientation control.
This is the unit-strength time convention used in the Letter.

At the unbiased endpoint the radial noise vanishes and
\begin{equation}
 \dot S=-2\ell^2\big[(1+\eta)S-(1-\eta)(1-S)\big],
 \qquad
 S_t^{\rm ub}=\frac{1-\eta}{2}
 +\left(S_0-\frac{1-\eta}{2}\right)e^{-4\ell^2t}.
 \label{eq:sm-ubbflow}
\end{equation}
Hence any inefficiency leaves the nonzero floor
\(S_{\rm stall}=(1-\eta)/2\).  At \(\eta=1\), this endpoint gives
the deterministic law \(S_t=S_0e^{-4\ell^2t}\).

\section{Exact QND channel and ambiguous records}
\label{sm:exact}

Assume \(0<\eta\leq1\), \(\ell>0\), and a mixed aligned initial state.
At \(x=\ell^2\), the state remains diagonal.  If
\(z=p_+-p_-\) and \(y=\operatorname{arctanh}z\), then
\(2y=\ln(p_+/p_-)\) is the accumulated evidence favoring one measurement
eigenstate over the other, and \(S=\tfrac12\operatorname{sech}^2y\).  The
QND dynamics becomes
\begin{equation}
 dy_t=\kappa\eta\tanh y_t\,dt+\sqrt{\kappa\eta}\,dW_t,
 \qquad \kappa=4\ell^2.
 \label{eq:sm-y}
\end{equation}
Its generator
\(\mathcal L_{\rm QND}=(\kappa\eta/2)\partial_y^2
+\kappa\eta\tanh y\,\partial_y\) is conjugate to free heat flow with a
constant spectral shift.  Explicitly,
\begin{equation}
 \E_{y_0}[f(y_t)]
 =\frac{e^{-\kappa\eta t/2}}{\cosh y_0}
 \big(G_{\kappa\eta t}*\cosh(\cdot)f(\cdot)\big)(y_0),
 \label{eq:sm-heat}
\end{equation}
where \(G_v\) is the centered Gaussian kernel of variance \(v\).
Equation~\eqref{eq:sm-heat} is the exact QND transition
kernel~\cite{Li2013,Jiang2020}.  Poltronieri Martins and Lima obtained its
perfect-record, maximally mixed specialization by exact path-integral
evaluation~\cite{Poltronieri2026}; their measurement-rate parameter
corresponds to the detected rate \(\kappa\eta\) here.  Related purification,
QND-collapse, and filter-stability results appear in
Refs.~\cite{MaassenKummerer2006,BauerBernard2011,BauerBenoistBernard2013,
BenoistPellegrini2014,vanHandel2009,AminiPellegriniRouchon2014}.

Three consequences are needed in the Letter.  First, the half-order identity
has a path-space interpretation.  Let the detected output satisfy
\[
 dY_t=2\sqrt{\eta}\Tr(L\rho_t)\,dt+dW_t,
\]
and let \(\mathsf P_\pm^{(t)}\) be its laws on \([0,t]\) conditioned on the
two QND eigenstates.  Relative to Wiener path measure \(\mathsf W^{(t)}\),
\[
 \frac{d\mathsf P_\pm^{(t)}}{d\mathsf W^{(t)}}
 =\exp\!\left(\pm2\sqrt{\eta}\ell Y_t-2\eta\ell^2t\right).
\]
Their Bhattacharyya coefficient, equivalently their Hellinger affinity, is
\[
 \mathcal B_t
 =\int\sqrt{d\mathsf P_+^{(t)}d\mathsf P_-^{(t)}}
 =e^{-2\eta\ell^2t}.
\]
Bayes' rule for arbitrary initial populations gives
\(\E_{\rm QND}\sqrt{S_t}/\sqrt{S_0}=\mathcal B_t\).  Thus taking
\(f=\operatorname{sech}\) makes the transformed weight constant and gives the
exact half-order identity
\begin{equation}
 \E_{\rm QND}\sqrt{S_t}=\sqrt{S_0}\,e^{-2\eta\ell^2t}.
 \label{eq:sm-qnd-root}
\end{equation}
Second, for \(f=\operatorname{sech}^{2\theta}\), the transformed weight is
\(\cosh^{1-2\theta}y\).  Gaussian asymptotics yield
\begin{equation}
 \Lambda_{\rm QND}(\theta,\eta)=
 \begin{cases}
 8\theta(1-\theta)\eta\ell^2,&0<\theta\leq\tfrac12,\\[2pt]
 2\eta\ell^2,&\theta\geq\tfrac12.
 \end{cases}
 \label{eq:sm-qnd-spectrum}
\end{equation}
For \(\theta>1/2\), the transformed weight is integrable and the spectral
edge supplies a \(t^{-1/2}\) prefactor.  At \(\theta=1/2\), the algebraic
scaling is \(t^0\).  For aligned QND and fixed \(\theta>1/2\), applying the
same edge law to the first and second moments gives
\(\operatorname{Var}(S_t^\theta)/\E[S_t^\theta]^2
\asymp t^{1/2}e^{2\eta\ell^2t}\).  Direct independent sampling at relative
root-mean-square error \(\epsilon\) therefore requires
\(N_{\rm traj}\asymp\epsilon^{-2}t^{1/2}e^{2\eta\ell^2t}\).  The tilted
measures constructed in Sec.~\ref{sm:alldfreeze} provide a natural
importance-sampling proposal, with likelihood-ratio reweighting returning the
physical moment.  Third, any fixed threshold \(0<s<1/2\) confines the
terminal evidence to a bounded interval, on which the heat kernel has the
same spectral-edge exponent.  Therefore
\begin{equation}
 \Pr_{\rm QND}(S_t\geq s)
 =e^{-2\eta\ell^2t+o(t)}.
 \label{eq:sm-qnd-survival}
\end{equation}
Equation~\eqref{eq:sm-qnd-survival} gives the fixed-QND threshold exponent.
Section~\ref{sm:alldfreeze} gives the protocol-uniform moment bound.

The same result can be read as a large-deviation saddle termination.  The
transition density associated with Eq.~\eqref{eq:sm-heat} is
\begin{equation}
 p_t(y\mid y_0)=
 \frac{e^{-v/2}\cosh y}{\cosh y_0}G_v(y-y_0),
 \qquad v=4\eta\ell^2t.
 \label{eq:sm-ydensity}
\end{equation}
For \(q_t=y_t/t\), its rate function is
\begin{equation}
 J(q)=2a+\frac{q^2}{8a}-|q|
 =\frac{(|q|-4a)^2}{8a},
 \qquad a=\eta\ell^2.
 \label{eq:sm-qrate}
\end{equation}
Because \(|-\ln S_t-2|y_t||\leq\ln2\), the contraction to
\(r_t=-t^{-1}\ln S_t\) gives
\begin{equation}
 I(r)=\frac{(r-8a)^2}{32a},
 \qquad r\geq0.
 \label{eq:sm-xrate}
\end{equation}
The Laplace principle then reads
\begin{equation}
 \Lambda_{\rm QND}(\theta,\eta)
 =\inf_{r\geq0}\{I(r)+\theta r\},
 \qquad r_\theta=\max\{8a(1-2\theta),0\}.
 \label{eq:sm-legendre}
\end{equation}
Below \(\theta=1/2\), a moving saddle selects trajectories with a positive
purification rate.  At \(\theta=1/2\), the saddle reaches \(r=0\).  Above this
value, the same rare records with \(O(1)\) net evidence dominate every higher
moment.
Equations~\eqref{eq:sm-qnd-spectrum} and
\eqref{eq:sm-qnd-survival} establish QND attainment and its physical
mechanism.  Protocol optimality requires the upper bound proved below.

\section{R\'enyi costs and the long-time spectrum}
\label{sm:renyi}

\subsection{Transfer of the long-time exponent}

For \(\alpha>0\), \(\alpha\ne1\), define the R\'enyi entropy
\begin{equation}
 H_\alpha(\rho)=\frac{1}{1-\alpha}\ln\Tr\rho^\alpha,
 \qquad q_\alpha=\min\{\alpha,1\}.
 \label{eq:sm-renyi-def}
\end{equation}
In every fixed finite dimension there are constants
\(0<c_{\alpha,d}\le C_{\alpha,d}<\infty\), independent of \(\rho\), such
that
\begin{equation}
 c_{\alpha,d}S^{q_\alpha}
 \le H_\alpha(\rho)
 \le C_{\alpha,d}S^{q_\alpha}.
 \label{eq:sm-renyi-compare}
\end{equation}
To see this, put \(\varepsilon=1-\lambda_{\max}(\rho)\).  Both
\(\varepsilon\) and \(S\) are comparable uniformly near the pure-state
boundary.  As \(\varepsilon\downarrow0\),
\(H_\alpha\asymp\varepsilon^\alpha\) for \(\alpha<1\) and
\(H_\alpha\asymp\varepsilon\) for \(\alpha>1\).  The ratio in
Eq.~\eqref{eq:sm-renyi-compare} is therefore uniformly bounded above and
below near the boundary.  For example, when \(\alpha<1\),
\(\varepsilon^\alpha\le\sum_{j>1}\lambda_j^\alpha
\le(d-1)^{1-\alpha}\varepsilon^\alpha\).  Compactness supplies global
constants away from that boundary.

Define the optimized qubit entropy exponent by
\begin{equation}
 \Gamma_\star^{(2)}(\alpha,\eta)
 =\liminf_{t\to\infty}-\frac1t
  \ln\inf_{\pi\in\Pi_t^{(2)}}\E_\pi[H_\alpha(\rho_t)].
\label{eq:sm-renyi-rate}
\end{equation}
Equation~\eqref{eq:sm-renyi-compare} and the qubit moment theorem give
\begin{equation}
 \Gamma_\star^{(2)}(\alpha,\eta)
 =\Lambda_\star^{(2)}(q_\alpha,\eta),
 \qquad \alpha\ne1.
 \label{eq:sm-renyi-transfer}
\end{equation}
At \(\alpha=1\), monotonicity of R\'enyi entropies and an elementary
finite-dimensional continuity bound give, for every \(\delta>0\),
\begin{equation}
 S\le H_1(\rho)\le C_{\delta,d}S^{1-\delta}.
 \label{eq:sm-vn-squeeze}
\end{equation}
Squeezing the exponential rate and using continuity of the moment spectrum
at \(\theta=1\) extends Eq.~\eqref{eq:sm-renyi-transfer} to
\(\alpha=1\).  Consequently the exact qubit result of the Letter becomes
\begin{equation}
 \Gamma_\star^{(2)}(\alpha,\eta)=
 \begin{cases}
 8\alpha(1-\alpha)\eta\ell^2,
 &0<\alpha<\tfrac12,\\[2pt]
 2\eta\ell^2,
 &\alpha\ge\tfrac12,\quad 0<\eta<1,\\[2pt]
 4q_\alpha\ell^2,
 &\alpha\ge\tfrac12,\quad \eta=1.
 \end{cases}
 \label{eq:sm-renyi-spectrum}
\end{equation}
Thus the entropy family has the same critical order \(\alpha=1/2\).  For
\(0<\alpha<1\), \(\alpha\) tilts the objective toward the same rare
high-impurity records, and at the critical order the dominant QND saddle
reaches the zero-evidence edge.  For \(\alpha\ge1\), the exponent-level tilt
saturates at \(q_\alpha=1\).

\section{Trace geometry and the finite-time moment envelope}
\label{sm:ddim}

\subsection{General-dimensional trace frame}

Let \(L\) be Hermitian with spectrum in an interval of width \(2\ell\), and
set \(\widetilde L=L-\Tr(\rho L)\openone\).  All orientation dependence of the
impurity generator is contained in
\begin{equation}
 T_1=\Tr(\widetilde L^2\rho^2),\qquad
 T_2=\Tr(\widetilde L\rho\widetilde L\rho),\qquad
 T_3=\Tr(\rho^2\widetilde L).
 \label{sm:ddim-frame}
\end{equation}
Their qubit values are Eqs.~\eqref{eq:sm-inv1}--\eqref{eq:sm-inv3}.  It\^o's
formula gives, in every finite dimension,
\begin{equation}
\begin{aligned}
 dS={}&-2B\,dt-4\sqrt\eta\,T_3\,dW_t,\\
 (dS)^2={}&16\eta T_3^2\,dt.
\end{aligned}
\label{sm:ddim-sde}
\end{equation}
\begin{equation}
 B=(1+\eta)T_2-(1-\eta)T_1
 =2\eta T_2-(1-\eta)\Delta,\qquad
 \Delta=T_1-T_2
 =\tfrac12\|[\widetilde L,\rho]\|_F^2\geq0.
 \label{eq:sm-Bdef}
\end{equation}
Unitary control preserves the state spectrum and changes the trace frame
through the predictable orientation presented to the next measurement.

For later use, applying It\^o's formula to \(\sqrt S\) gives
\begin{equation}
 -\frac{\mathcal A\sqrt S}{\sqrt S}
 =\frac{B}{S}+2\eta\frac{T_3^2}{S^2}.
 \label{sm:ddim-sqrtgen}
\end{equation}
The desired pointwise half-moment bound is therefore equivalent, for \(S>0\),
to the master inequality
\begin{equation*}
 B+2\eta\frac{T_3^2}{S}\leq2\eta\ell^2S.
 \tag{\(C_{\sqrt S}\)}
 \label{sm:csqrt-target}
\end{equation*}
We prove it from two sharp spectral inequalities.

\subsection{A bound on radial innovation}

\textbf{Lemma A.}
\label{sm:lemA}
For every state and every centered Hermitian \(\widetilde L\) of spectral
half-spread \(\ell\),
\begin{equation}
 |T_3|=|\Tr(\rho^2\widetilde L)|\leq\ell S.
 \label{eq:sm-lemA}
\end{equation}
\emph{Proof.}
Work in the eigenbasis
\(\rho=\operatorname{diag}(\lambda_1,\ldots,\lambda_d)\), and let
\(m_i=\widetilde L_{ii}\).  The diagonal entries lie in some interval of
width \(2\ell\), and centering gives \(\sum_i\lambda_i m_i=0\).  Write
\(\widetilde m_i=m_i-(u+\ell)\in[-\ell,\ell]\) and
\(s=\sum_i\lambda_i\widetilde m_i=-(u+\ell)\).  With
\(r_2=\sum_i\lambda_i^2\), this gives
\(T_3=\sum_i\lambda_i(\lambda_i-r_2)\widetilde m_i\).
Pointwise, \(|\lambda_i-r_2|\leq1-\lambda_i\).  The negative-sign case
follows from \(r_2\leq1\), while the positive-sign case follows from
\(\lambda_i^2\leq r_2\) and
\(2\lambda_i\leq1+\lambda_i^2\).  Hence
\(|T_3|\leq\ell\sum_i\lambda_i(1-\lambda_i)
=\ell(1-\Tr\rho^2)=\ell S\).
\(\square\)

\subsection{Quartic inequality and equality geometry}
\label{sm:ct2}

\textbf{Theorem.}
Let \(P_\rho\) be the support projector of \(\rho\), with the width of
\(P_\rho L P_\rho\) understood on \(\operatorname{supp}\rho\).  Then
\begin{equation}
 T_2S+T_3^2\leq\ell^2S^2.
 \label{eq:sm-ct2}
\end{equation}
If \(S>0\) and \(\ell>0\), equality holds precisely when
\begin{equation}
 \operatorname{rank}\rho=2,\qquad
 \operatorname{width}(P_\rho L P_\rho)=2\ell.
 \label{eq:sm-ct2-equality}
\end{equation}
For \(S=0\) or \(\ell=0\), the equality follows directly.

\emph{Proof.}
In an eigenbasis of \(\rho\), put
\(N_\tau=\widetilde L+\tau\openone\).  Since
\(\Tr(\rho\widetilde L)=0\),
\begin{align}
 T_2+2\tau T_3-\tau^2S
 &=2\sum_{i<j}\lambda_i\lambda_j
 \left(|(N_\tau)_{ij}|^2
 -(N_\tau)_{ii}(N_\tau)_{jj}\right)\nonumber\\
 &=2\sum_{i<j}\lambda_i\lambda_j[-\det B_{ij}(\tau)],
 \label{eq:sm-pairdet}
\end{align}
where \(B_{ij}(\tau)\) is the corresponding \(2\times2\) principal
compression.  Write its eigenvalues as \(c_{ij}\pm d_{ij}\).
Every principal compression has half-width \(0\leq d_{ij}\leq\ell\), and
\[
 -\det B_{ij}(\tau)=d_{ij}^2-c_{ij}^2\leq\ell^2.
\]
Since \(S=2\sum_{i<j}\lambda_i\lambda_j\), Eq.~\eqref{eq:sm-pairdet} gives
\[
 T_2+2\tau T_3-\tau^2S\leq\ell^2S.
\]
For \(S>0\), maximizing at \(\tau_\star=T_3/S\) proves
Eq.~\eqref{eq:sm-ct2}.

For equality, every positive-weight pair compression at \(\tau_\star\) must
have eigenvalues exactly \(\pm\ell\).  If three indices \(i,j,k\) belonged to
the support, the three pairwise trace conditions would force the corresponding
diagonal entries of \(N_{\tau_\star}\) to vanish, while the determinant
conditions would give all three off-diagonal moduli equal to \(\ell\).
The \(3\times3\) compression \(C\) would then satisfy
\(\Tr C=0\) and \(\Tr C^2=6\ell^2\).  A three-point spectrum of width at
most \(2\ell\) obeys
\[
 \frac13\Tr C^2\leq\frac{\operatorname{width}(C)^2}{4}\leq\ell^2,
\]
a contradiction.  Thus equality requires rank two.

Let the two nonzero eigenvalues be \(p,q\), and let \(a,b\) be the
corresponding diagonal entries of \(\widetilde L\).  Centering gives
\(pa+qb=0\), while \(S=2pq\), so
\(\tau_\star=T_3/S=-(a+b)/2\).  The support compression of
\(N_{\tau_\star}\) is therefore traceless and has eigenvalues
\(\pm\ell\) exactly when the compression of \(L\) has width \(2\ell\).
This proves necessity and sufficiency. \(\square\)

\begin{figure}[t]
\centering
\includegraphics[width=0.92\textwidth]{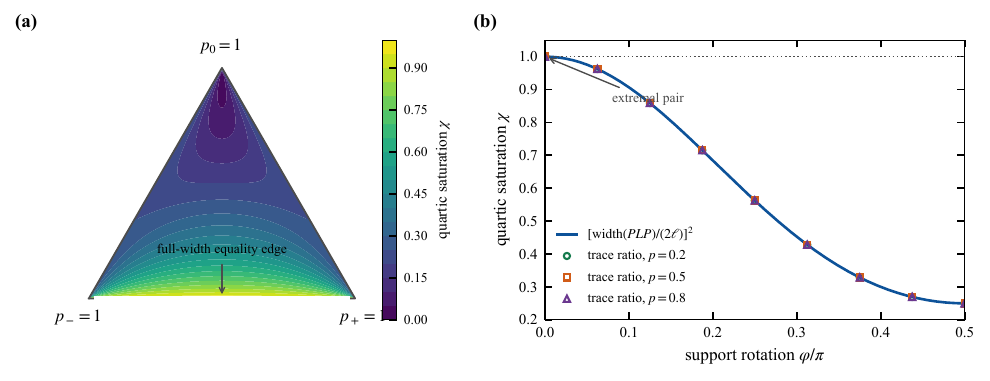}
\caption{Equality geometry for
\(L=\ell\,\operatorname{diag}(-1,0,1)\).
(a) On the commuting diagonal slice, the color is
\(\chi=(T_2S+T_3^2)/(\ell^2S^2)\).  Among mixed states, the extremal edge
\(p_0=0\) saturates, while the other rank-two edges have \(\chi=1/4\).
(b) For a rank-two support spanned by
\(\lvert-\rangle\) and
\(\cos\varphi\lvert+\rangle+\sin\varphi\lvert0\rangle\),
\(\chi=[(1+\cos^2\varphi)/2]^2\), independently of the two nonzero
eigenvalues.  The curve measures the width of the support compression.}
\label{fig:sm-equality}
\end{figure}

For a qubit with \(S>0\), the support is the whole two-dimensional space, so
Eq.~\eqref{eq:sm-ct2} is an identity for every orientation.  In higher
dimension, equality additionally requires the support to span the full
spectral width, as Fig.~\ref{fig:sm-equality} illustrates.

\subsection{Master inequality, equality, and finite-time regularity}
\label{sm:lemR}

Define
\[
 P=(T_1-T_2)S,\qquad
 Q=(T_1+T_2)S+2T_3^2-2\ell^2S^2.
\]
Then \(P=\tfrac12\|[\widetilde L,\rho]\|_F^2S\geq0\).  After multiplying
Eq.~\eqref{sm:csqrt-target} by \(S\), its excess is
\[
 -P+\eta Q.
\]
Moreover, \(Q\leq P\) is exactly
\(T_2S+T_3^2\leq\ell^2S^2\).  For \(Q\geq0\),
\(\eta Q\leq\eta P\leq P\).  For \(Q<0\),
\(\eta Q\leq0\leq P\).  Thus the quartic theorem proves
\begin{equation}
 (1+\eta)T_2-(1-\eta)T_1
 +2\eta\frac{T_3^2}{S}
 \leq2\eta\ell^2S
 \label{sm:csqrt}
\end{equation}
for every \(0\leq\eta\leq1\).

The equality conditions are also explicit.  At \(\eta=1\), equality in
Eq.~\eqref{sm:csqrt} is precisely equality in the quartic theorem.  For
\(0<\eta<1\), equality in \(\eta Q\leq P\) requires \(P=Q=0\).
For \(S>0\) and \(\ell>0\), the full half-moment generator therefore
saturates pointwise precisely when the state has an extremal rank-two support
and \([L,\rho]=0\).  Equality in a finite-time expectation bound requires the
same conditions along the trajectory for
\(dt\otimes d\mathbb P\)-almost every point.

Before applying singular functions of \(S\), we record the required
regularity.  Put
\[
 X=\tfrac12\ln S,\qquad
 \nu=-2\sqrt\eta\,\frac{T_3}{S}.
\]
For \(n>S_0^{-1}\), stop at
\(\tau_n=\inf\{t:S_t\leq n^{-1}\}\).  Up to these stopping times,
Eqs.~\eqref{sm:ddim-sde} and \eqref{sm:csqrt} give
\begin{equation}
 dX=-\left(\frac BS+\nu^2\right)dt+\nu\,dW_t,
 \qquad
 \nu^2\leq4\eta\ell^2,\qquad
 \frac BS+\nu^2\leq4\eta\ell^2.
 \label{eq:sm-logS}
\end{equation}
Extend \(\nu\) predictably by zero at \(S=0\), and let
\(N_t=\int_0^t\nu_u\,dW_u\).  Fix a terminal \(t<\infty\).  For
\(0\leq v\leq t\),
\[
 X_{v\wedge\tau_n}\geq X_0-4\eta\ell^2t
 -\sup_{u\leq t}|N_u|.
\]
The right side is finite almost surely and independent of \(n\).  If a path
reached \(S=0\) by time \(t\), continuity would give \(\tau_n\leq t\) for
every sufficiently large \(n\), contradicting the common lower bound because
\(X_{\tau_n}=-(1/2)\ln n\).  Hence every mixed initial state, including a
rank-deficient one, satisfies \(S_t>0\) at every finite time almost surely.

\subsection{Finite-time moment envelope at the critical order}
\label{sm:ddimmoment}
\label{sm:moments}

For every fixed \(\theta>0\), It\^o's formula gives
\begin{equation}
 -\frac{\mathcal A S^\theta}{S^\theta}
 =2\theta\frac{B}{S}
 +8\theta(1-\theta)\eta\frac{T_3^2}{S^2}.
 \label{eq:sm-momgen}
\end{equation}
Using Eq.~\eqref{sm:csqrt} to bound \(B\) reduces this to
\[
 -\frac{\mathcal A S^\theta}{S^\theta}
 \leq4\theta\eta\ell^2
 +4\theta(1-2\theta)\eta\frac{T_3^2}{S^2}.
\]
For \(0<\theta\leq1/2\), Lemma A bounds the residual term.  For
\(\theta\geq1/2\), that term is nonpositive.  Therefore
\begin{equation}
\begin{aligned}
 -\frac{\mathcal A S^\theta}{S^\theta}
 &\leq c(\theta)\eta\ell^2,\\
 c(\theta)&=
 \begin{cases}
 8\theta(1-\theta),&0<\theta\leq\tfrac12,\\[2pt]
 4\theta,&\theta\geq\tfrac12,
 \end{cases}\\
 \E_\pi[S_t^\theta]
 &\geq S_0^\theta e^{-c(\theta)\eta\ell^2t}.
\end{aligned}
\label{sm:reduction}
\end{equation}
The last line follows from the exponential submartingale
\(e^{c(\theta)\eta\ell^2t}S_t^\theta\), stopped away from \(S=0\), followed
by the finite-time regularity above and bounded convergence.  The bound holds
for every finite dimension, every \(t\geq0\), every fixed finite
\(\theta>0\), and every admissible predictable basis policy.

\subsubsection{Qubit half-moment identity}
\label{sm:sqrt}

The qubit half-moment equality also follows directly.
With \(h(x,S)=g(x,S)/S\), Eq.~\eqref{eq:sm-sde} gives
\begin{equation}
 -\frac{\mathcal A\sqrt S}{\sqrt S}
 =h(x,S)+2\eta x\beta ,
 \qquad
 \partial_x[h+2\eta x\beta]
 =\frac{(1-\eta)(1-2S)}{S}.
 \label{eq:sm-sqrtgen}
\end{equation}
The maximum is at the aligned endpoint, where
\begin{equation}
 [h+2\eta x\beta]_{x=\ell^2}
 =4\eta\ell^2S+2\eta\ell^2(1-2S)
 =2\eta\ell^2.
 \label{eq:sm-pivot}
\end{equation}
Thus \(\E\sqrt{S_t}\geq\sqrt{S_0}e^{-2\eta\ell^2t}\).  For
\(0<\eta<1\), pointwise equality away from the maximally mixed qubit requires
the QND endpoint.  At \(S=1/2\), every basis commutes with the state and ties
instantaneously.  At \(\eta=1\), every mixed-qubit orientation is a pointwise
equality case.  Every admissible qubit basis history then satisfies
\[
 \E_\pi\sqrt{\det\rho_t}
 =\sqrt{\det\rho_0}\,e^{-2\ell^2t}.
\]
In any dimension, a commuting mixture of the two extremal eigenvectors of
\(L\) remains supported on an invariant two-dimensional subspace and saturates
the half-moment bound for all time.  Away from the critical order, QND states
supported on this subspace attain the low-order finite-time envelope at the
exponential scale.

\section{All-dimensional inefficient ceiling}
\label{sm:alldfreeze}

We now establish a horizon-uniform upper bound in the high-order regime.  Fix
a finite dimension, a mixed initial state \(S_0>0\), an efficiency
\(0<\eta<1\), and a fixed finite \(\theta>0\).  Then
\begin{equation}
 \limsup_{t\to\infty}
 -\frac1t\ln V_\theta^{(d)}(t,\eta)\leq2\eta\ell^2.
 \label{eq:sm-alld-ceiling}
\end{equation}
Together with Eq.~\eqref{sm:reduction}, this gives the state-space-uniform
ceiling
\begin{equation}
 \Lambda_\star^{(d)}(\theta,\eta)\leq
 \begin{cases}
 8\theta(1-\theta)\eta\ell^2,
     &0<\theta\leq\tfrac12,\\[2pt]
 2\eta\ell^2,&\theta\geq\tfrac12,
 \end{cases}
 \qquad 0<\eta<1.
 \label{eq:sm-alld-piecewise}
\end{equation}

\subsection{Near-pure rigidity}

The first ingredient bounds radial drift when the state is nearly pure.
Whenever \(S\leq1/4\),
\begin{equation}
 T_2\leq12\ell^2S^2+8S\Delta ,
 \qquad \Delta=T_1-T_2.
 \label{eq:sm-nearpure}
\end{equation}
To prove it, diagonalize
\(\rho=\operatorname{diag}(p_1,\ldots,p_d)\), with \(p_1=\max_i p_i\), and
write \(\varepsilon=1-p_1\).  Since
\(\sum_i p_i^2\leq p_1\), one has \(\varepsilon\leq S\), and
\(S\leq1/4\) implies \(p_1\geq3/4\).  Shift \(L\) by the midpoint of its
spectral interval so that \(\|L\|\leq\ell\), and set
\(m_i=\widetilde L_{ii}\).  Centering gives
\[
 m_1=\sum_{j>1}p_j(L_{11}-L_{jj}),
\]
so \(|m_1|\leq2\ell\varepsilon\), while
\(|m_j|\leq(5/2)\ell\) for \(j>1\).  In this basis,
\begin{align}
 T_2&=\sum_i p_i^2m_i^2
 +2\sum_{i<j}p_ip_j|L_{ij}|^2,
 \label{eq:sm-t2split}\\
 \Delta&=\sum_{i<j}(p_i-p_j)^2|L_{ij}|^2.
 \label{eq:sm-deltasplit}
\end{align}
The diagonal term and pairs with \(i,j>1\) are at most
\[
 \left(4+\frac{25}{4}+1\right)\ell^2\varepsilon^2
 \leq12\ell^2S^2.
\]
For every dominant pair \((1,j)\),
\(2p_1p_j\leq2\varepsilon\leq
8\varepsilon(p_1-p_j)^2\), because \(p_1-p_j\geq1/2\).  Summing these
pairs and using \(\varepsilon\leq S\) proves
Eq.~\eqref{eq:sm-nearpure}.

Since \(B=2\eta T_2-(1-\eta)\Delta\),
Eq.~\eqref{eq:sm-nearpure} gives
\[
 B\leq24\eta\ell^2S^2+
 [16\eta S-(1-\eta)]\Delta.
\]
Therefore, with
\[
 s_\eta=\min\left\{\frac14,\frac{1-\eta}{16\eta}\right\},
\]
the protocol-independent near-pure estimate is
\begin{equation}
 S\leq s_\eta\quad\Longrightarrow\quad
 B\leq24\eta\ell^2S^2.
 \label{eq:sm-Blocal}
\end{equation}

\subsection{Corridor change of measure}

Use \(X\) and \(\nu\) from Eq.~\eqref{eq:sm-logS}.  Fix a dimensionless
horizon \(t\), an arbitrary rule \(\pi\in\Pi_t^{(d)}\), and
\(0<s<s_\eta\), and let
\(\chi_u=\mathbf1_{\{S_{u-}\leq s\}}\).  Continuity permits \(S_u\) in all
integrals, while the left-limit notation makes \(\chi\) predictable.  For
\(0\leq v\leq t\), define
\begin{equation}
 Z_v=\exp\left\{
 \int_0^v\chi_u\nu_u\,dW_u
 -\frac12\int_0^v\chi_u\nu_u^2\,du
 \right\}.
 \label{eq:sm-girsanov-density}
\end{equation}
The integrand is bounded, so Novikov's condition holds.  Under the
horizon-indexed measure \(d\mathbb Q^{(t)}=Z_t\,d\mathbb P\),
\[
 W_v^{(t)}=W_v-\int_0^v\chi_u\nu_u\,du
\]
is Brownian motion on \([0,t]\).  Equation~\eqref{eq:sm-logS} becomes
\begin{equation}
 dX=\left[-\frac BS-(1-\chi)\nu^2\right]du
 +\nu\,dW_u^{(t)}.
 \label{eq:sm-logS-Q}
\end{equation}
The dependence of \(Z\), \(\mathbb Q^{(t)}\), and \(W^{(t)}\) on the selected
rule \(\pi\) is suppressed in the notation.  The auxiliary measure may depend
on \((t,\pi)\), whereas every constant in the physical-measure bound below is
uniform over both.
The measures are equivalent and leave the filtration unchanged, so the
selected rule remains predictable under \(\mathbb Q^{(t)}\).
Inside this region its drift is at least
\(-\kappa_s\), where \(\kappa_s=24\eta\ell^2s\).

Set
\[
 \mathcal M_v^{(t)}
 =\int_0^v\chi_u\nu_u\,dW_u^{(t)},\qquad
 \mathcal V_v=\int_0^v\chi_u\nu_u^2\,du
 \leq v_{\max}v,\qquad v_{\max}=4\eta\ell^2.
\]
Let \(c=\tfrac12\ln s\) and \(c_0=\min\{X_0,c\}\).  Integrating after the
last crossing of \(c\), or from zero for trajectories that stay below \(c\),
gives
\begin{equation}
 X_t\geq c_0-\kappa_st
 -2\sup_{v\leq t}|\mathcal M_v^{(t)}|.
 \label{eq:sm-last-excursion}
\end{equation}
For \(\gamma>0\), put
\(F_t=\{\sup_{v\leq t}|\mathcal M_v^{(t)}|\leq\gamma t\}\).  The
exponential-martingale maximal inequality yields
\begin{equation}
 \mathbb Q^{(t)}(F_t^c)
 \leq2\exp\left[-\frac{\gamma^2t}{8\eta\ell^2}\right].
 \label{eq:sm-Mmax}
\end{equation}
Written in terms of \(W^{(t)}\), the reverse density is
\begin{equation}
 \left.\frac{d\mathbb P}{d\mathbb Q^{(t)}}\right|_{\mathcal F_t}
 =Z_t^{-1}
 =\exp\left[-\mathcal M_t^{(t)}-\frac12\mathcal V_t\right].
 \label{eq:sm-reverse-density}
\end{equation}
On \(F_t\), Eqs.~\eqref{eq:sm-last-excursion}--\eqref{eq:sm-reverse-density}
therefore imply, for all sufficiently large \(t\),
\begin{align}
 \E_{\mathbb P}[S_t^\theta]
 &\geq
 \left(1-2e^{-\gamma^2t/(8\eta\ell^2)}\right)
 \min\{S_0,s\}^{\theta}\nonumber\\
 &\quad\times
 \exp\left[-\{2\eta\ell^2
 +48\theta\eta\ell^2s+(1+4\theta)\gamma\}t\right].
 \label{eq:sm-corridor-bound}
\end{align}
The right-hand side of Eq.~\eqref{eq:sm-corridor-bound} and the minimum
\(t\) for which that bound holds are independent of
\(\pi\in\Pi_t^{(d)}\).  We may therefore take the policy infimum first, then
let \(t\to\infty\) at fixed \(s,\gamma,\theta\), and finally let
\(\gamma\downarrow0\) and \(s\downarrow0\).  This proves
Eq.~\eqref{eq:sm-alld-ceiling}.  The measures \(\mathbb Q^{(t)}\) are
constructed separately for each horizon and selected rule; every displayed
bound remains an inequality under the physical measure.

\subsection{Higher-dimensional radial geometry}

The change of measure is needed because higher dimensions admit directions
with vanishing radial diffusion and nonzero inward drift.  For example, take
\[
 \rho=\operatorname{diag}(1-2\varepsilon,\varepsilon,\varepsilon),
 \qquad
 L=\begin{pmatrix}
 0&0&0\\
 0&0&\ell\\
 0&\ell&0
 \end{pmatrix}.
\]
Then \(T_3=0\) and \(T_1=T_2=2\ell^2\varepsilon^2\), so
\(B=4\eta\ell^2\varepsilon^2>0\).  The impurity therefore decreases with
zero instantaneous radial diffusion.  Equation~\eqref{eq:sm-Blocal}
quantifies the drift at order \(O(S^2)\), which gives zero local exponential
rate.  This qutrit direction displays zero radial diffusion together with
\(O(S^2)\) inward drift.

\subsection{Sharpness on an embedded qubit face}
\label{sm:embed}

Let \(\lvert+\rangle,\lvert-\rangle\) be eigenvectors of \(L\) at its two
spectral endpoints, and initialize
\[
 \rho_0=p\lvert+\rangle\langle+\rvert
 +(1-p)\lvert-\rangle\langle-\rvert,\qquad 0<p<1.
\]
Fixed QND monitoring leaves this support invariant and is exactly the qubit
channel of Sec.~\ref{sm:exact}.  Its spectrum
Eq.~\eqref{eq:sm-qnd-spectrum} attains both cases of
Eq.~\eqref{eq:sm-alld-piecewise}.  Thus the ceiling is sharp over the allowed
state space in every finite dimension.  This construction establishes
attainment for commuting states supported on the extremal two-dimensional
subspace.

\section{Qubit fixed-threshold certificates}
\label{sm:frozen}

For a qubit and \(0<\eta<1\), we now strengthen the moment upper bound with a
family of event-level certificates.  At every fixed
\(0<\omega\leq\sqrt{1-\eta}\), define the certificate threshold by
\[
 s_Y=\tfrac12\operatorname{sech}^2Y,\qquad
 Y=\frac{\pi}{2\omega}.
\]
Recovery of the limiting ceiling \(2\eta\ell^2\) requires the ordered limits
\(t\to\infty\) and then \(\omega\downarrow0\).  Every fixed threshold has
the rate specified below.

For an arbitrary predictable qubit orientation
\(x_t\in[0,\ell^2]\), apply It\^o's formula to the radial state coordinate
\(b=\operatorname{arctanh}\sqrt{1-2S}\).  In the interior \(b>0\),
\begin{equation}
 db=
 \frac{g(x,S)-2\eta xS(1-3\beta)}
 {S\sqrt\beta}\,dt
 +2\sqrt{\eta x}\,dW_t.
 \label{eq:sm-bdrift}
\end{equation}
At \(b=0\) the radial variable is reflected.  The even verification
function below has zero derivative there, so its local-time term vanishes and
the generator extends by continuity.  Define, for
\(S\in[s_Y,1/2]\),
\begin{equation}
 \psi_\omega(S)
 =\operatorname{sech}b\,\cos(\omega b),
 \qquad
 \lambda_\omega=2\eta\ell^2(1+\omega^2).
 \label{eq:sm-psi}
\end{equation}
It obeys \(0\leq\psi_\omega\leq1\),
\(\psi_\omega(s_Y)=0\), and the pointwise inequality
\begin{equation}
 \mathcal A_x\psi_\omega\geq-\lambda_\omega\psi_\omega
 \qquad
 (0\leq x\leq\ell^2,\ s_Y\leq S\leq\tfrac12).
 \label{eq:sm-genineq}
\end{equation}

The drift and squared diffusion in Eq.~\eqref{eq:sm-bdrift} are affine in
\(x\), so it is enough to check the endpoints.  At \(x=\ell^2\), the
ground-state transform of Sec.~\ref{sm:exact} gives
\[
 \mathcal A_{\ell^2}[\operatorname{sech}b\,v(b)]
 =2\eta\ell^2\operatorname{sech}b\,[v''(b)-v(b)].
\]
With \(v(b)=\cos(\omega b)\), this is the exact eigenrelation
\(\mathcal A_{\ell^2}\psi_\omega=-\lambda_\omega\psi_\omega\).

At \(x=0\), the motion is deterministic with drift
\[
 D_0(b)
 =2\ell^2\,\frac{1-(1-\eta)\cosh^2b}{\tanh b}
 =2\ell^2\,\frac{\eta-u}{\tanh b},
 \qquad u=(1-\eta)\sinh^2b.
\]
If \(D_0\leq0\), then \(\psi_\omega'(b)\leq0\), and
\(\mathcal A_0\psi_\omega=D_0\psi_\omega'\geq0\).  If \(D_0>0\), then
\(0\leq u<\eta\).  The choice
\(\omega^2\leq1-\eta\) also ensures \(0\leq\omega b<\pi/2\) throughout
this region because
\(\operatorname{arctanh}z\leq z/\sqrt{1-z^2}\) for \(0\leq z<1\), and
hence \(\sqrt{1-\eta}\operatorname{arctanh}\sqrt\eta\leq\sqrt\eta<1\).
After substituting
\(D_0\) and \(\psi_\omega'\), Eq.~\eqref{eq:sm-genineq} is equivalent to
\begin{equation}
 (u+\eta\omega^2)\cos(\omega b)
 \geq(\eta-u)\omega\sin(\omega b)\coth b.
 \label{eq:sm-star}
\end{equation}
Put \(w=\omega^2\).  Using
\(\sin(\omega b)\leq\omega b\),
\(\coth b\leq b^{-1}+b/3\), and
\(\cos(\omega b)\geq1-wb^2/2\), define the exact residual
\[
 \Delta_{\eta,\omega}(b)
 :=(u+\eta w)\cos(\omega b)
 - (\eta-u)\omega\sin(\omega b)\coth b.
\]
It obeys
\begin{align}
 \Delta_{\eta,\omega}(b)
 &\geq(u+\eta w)\left(1-\frac{wb^2}{2}\right)
 -(\eta-u)w\left(1+\frac{b^2}{3}\right)\nonumber\\
 &=u(1+w)-\frac{wb^2}{6}[u+2\eta+3\eta w].
 \label{eq:sm-star-chain}
\end{align}
Since \(u\leq\eta\) and
\(u=(1-\eta)\sinh^2b\geq(1-\eta)b^2\),
\[
 \Delta_{\eta,\omega}(b)
 \geq b^2(1+w)\left[(1-\eta)-\frac{\eta w}{2}\right]\geq0.
\]
This proves Eq.~\eqref{eq:sm-genineq} at both endpoints and hence for every
orientation.  Equivalently, with \(u_x=x/\ell^2\),
\begin{equation}
 \frac{\mathcal A_x\psi_\omega+\lambda_\omega\psi_\omega}
 {2\ell^2\operatorname{sech}b}
 =(1-u_x)\Delta_{\eta,\omega}(b)\geq0.
 \label{eq:sm-residual-factorization}
\end{equation}

Choose \(\omega\) small enough that \(s_Y<S_0\), and let
\(\tau_Y=\inf\{t:S_t\leq s_Y\}\).  The stopped process
\(e^{\lambda_\omega(t\wedge\tau_Y)}
\psi_\omega(S_{t\wedge\tau_Y})\) is a nonnegative local submartingale.
Localization and bounded convergence give
\begin{equation}
 \psi_\omega(S_0)
 \leq e^{\lambda_\omega t}\Pr(\tau_Y>t).
 \label{eq:sm-survival}
\end{equation}
On \(\{\tau_Y>t\}\), one has \(S_t\geq s_Y\), and therefore, for every
\(\theta>0\),
\begin{equation}
 \E[S_t^\theta]\geq
 s_Y^\theta\psi_\omega(S_0)
 e^{-2\eta\ell^2(1+\omega^2)t}.
 \label{eq:sm-frozenbound}
\end{equation}

\begin{figure}[t]
\centering
\includegraphics[width=0.92\textwidth]{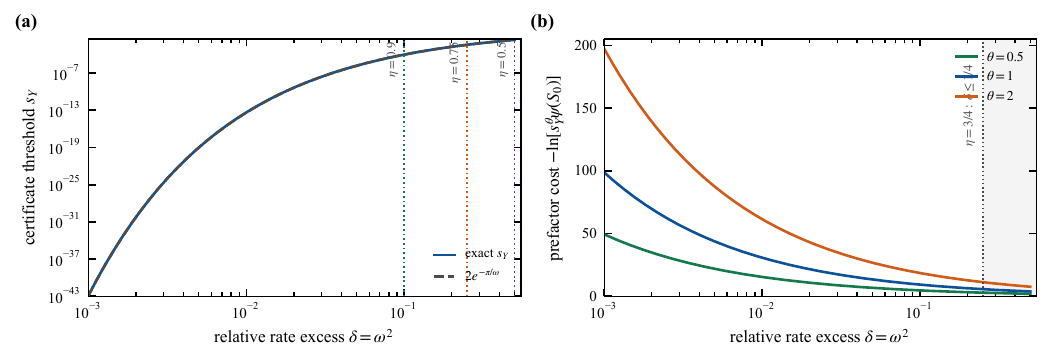}
\caption{Cost of approaching the frozen rate.
(a) The certificate threshold and its small-\(\delta\) asymptote.  Vertical
lines mark the admissible endpoints \(\delta=1-\eta\).
(b) For \(S_0=0.455\), the prefactor cost
\(-\ln[s_Y^\theta\psi_\omega(S_0)]\) diverges as
\(\delta\downarrow0\).  Every fixed certificate has
\(\lambda_\omega>2\eta\ell^2\).  The ordered limits
\(t\to\infty\) followed by \(\omega\downarrow0\) recover the sharp moment
ceiling.}
\label{fig:sm-threshold-tradeoff}
\end{figure}

The limiting construction has a nonuniform prefactor.  If
\(\delta_\omega=\lambda_\omega/(2\eta\ell^2)-1=\omega^2\), then
\begin{equation}
 s_Y=\frac12\operatorname{sech}^2
 \left(\frac{\pi}{2\sqrt{\delta_\omega}}\right),
 \qquad
 0<\delta_\omega\leq1-\eta ,
 \label{eq:sm-threshold-tradeoff}
\end{equation}
and \(s_Y\sim2e^{-\pi/\sqrt{\delta_\omega}}\) as
\(\delta_\omega\downarrow0\).  For fixed mixed \(S_0\),
\[
 -\ln[s_Y^\theta\psi_\omega(S_0)]
 =\frac{\pi\theta}{\sqrt{\delta_\omega}}+O(1).
\]
Figure~\ref{fig:sm-threshold-tradeoff} displays the tradeoff between rate and
prefactor.

For each such fixed \(\omega\), the prefactor in
Eq.~\eqref{eq:sm-frozenbound} is positive and independent of \(t\).
Taking \(t\to\infty\) first and then \(\omega\downarrow0\) recovers the
qubit high-order ceiling \(2\eta\ell^2\).  This event-level construction
strengthens the qubit upper bound, and Sec.~\ref{sm:alldfreeze} extends the
same ceiling to every finite dimension.

\subsection{Qubit optimum over time-horizon-dependent policies}

The bounds above determine the optimized exponential rate for rule classes
chosen separately at each time horizon.  For \(0<\theta\leq1/2\),
Eq.~\eqref{sm:reduction} is uniform over \(\Pi_t^{(2)}\), and QND has the
matching exponent.  For \(\eta=1\) and \(\theta>1/2\), the same envelope is
saturated at every time by the deterministic unbiased flow.  For
\(0<\eta<1\) and \(\theta>1/2\), Eq.~\eqref{eq:sm-frozenbound} remains valid
after taking the policy infimum because its prefactor is independent of the
rule and time horizon.  Consequently the full limit exists:
\begin{equation}
 \lim_{t\to\infty}-\frac1t\ln V_\theta^{(2)}(t,\eta)
 =\Lambda_\star^{(2)}(\theta,\eta)
 =\begin{cases}
 8\theta(1-\theta)\eta\ell^2,
 &0<\theta<\tfrac12,\quad 0<\eta\leq1,\\[2pt]
 2\eta\ell^2,
 &\theta\geq\tfrac12,\quad 0<\eta<1,\\[2pt]
 4\theta\ell^2,
 &\theta\geq\tfrac12,\quad \eta=1.
 \end{cases}
 \label{eq:sm-horizon-qubit}
\end{equation}
The theorem fixes the time-horizon-uniform exponential rate.  Time-horizon-dependent
rules may still change finite-time values and subexponential factors.

\section{Full-rank determinant separation}
\label{sm:detlaw}

The uniform ceiling is attained on an extremal rank-two boundary.  A
determinant estimate yields a stricter bound for generic full-rank interior
states.  Define
\begin{equation}
 \mu=\frac{\Tr L}{d},\qquad
 \widehat L=L-\mu\openone,\qquad
 V_L=\frac1d\Tr\widehat L^2,\qquad
 R=(\det\rho)^{1/d}.
 \label{eq:sm-VL}
\end{equation}
The spectral variance \(V_L\) is invariant under every allowed orientation.
For a minimal purification, \(d\,R\) is the G-concurrence of the Schmidt
spectrum~\cite{Gour2005,Konrad2008,Tiersch2008,Gour2010}.  The
determinant-root theorem below governs its controlled inefficient diffusion.

\textbf{Determinant-root theorem.}
For every full-rank initial state, every finite \(t\), every
\(0\leq\eta\leq1\), and every admissible predictable policy,
\begin{equation}
 \E_\pi[R_t]\geq R_0e^{-2\eta V_Lt}.
 \label{eq:sm-detroot-bound}
\end{equation}
At \(\eta=1\), equality holds for every policy.  For \(0\leq\eta<1\),
equality holds exactly when \([\rho_u,L_u]=0\) for
\(du\otimes d\mathbb P\)-almost every \((u,\omega)\) up to the horizon.
Fixed QND monitoring from a commuting state is one such policy.

\subsection{Matrix It\^o calculation}

The stochastic master equation is invariant under scalar shifts of \(L\).
We may therefore take \(\mu=0\) and set
\[
 m=\Tr(\rho L),\qquad
 G=\sqrt\eta(L\rho+\rho L-2m\rho),\qquad
 \mathcal Q=\Tr(\rho^{-1}L\rho L).
\]
For \(d\rho=\mathcal D[L]\rho\,dt+G\,dW_t\), matrix It\^o calculus gives
\begin{equation}
 d\ln\det\rho
 =\Tr(\rho^{-1}d\rho)
 -\frac12\Tr(\rho^{-1}G\rho^{-1}G)\,dt.
 \label{eq:sm-logdet-ito}
\end{equation}
The required traces are
\begin{align}
 \Tr[\rho^{-1}\mathcal D[L]\rho]
 &=\mathcal Q-\Tr L^2,
 \label{eq:sm-logdet-drift1}\\
 \Tr(\rho^{-1}G)
 &=-2d\sqrt\eta\,m,
 \label{eq:sm-logdet-noise}\\
 \Tr[(\rho^{-1}G)^2]
 &=\eta[2\Tr L^2+2\mathcal Q+4dm^2].
 \label{eq:sm-logdet-quad}
\end{align}
Substitution yields
\begin{equation}
 d\ln\det\rho
 =[(1-\eta)\mathcal Q-(1+\eta)\Tr L^2-2\eta dm^2]dt
 -2d\sqrt\eta\,m\,dW_t.
 \label{eq:sm-logdet}
\end{equation}
Applying scalar It\^o calculus to \(R\) cancels the \(m^2\) term and yields
\begin{align}
 \frac{dR}{R}
 &=\left[-2\eta V_L+\frac{1-\eta}{d}K(\rho,L)\right]dt
 +2\sqrt\eta(\mu-\Tr\rho L)\,dW_t,
 \label{eq:sm-detroot-sde}\\
 K(\rho,L)
 &:=\Tr(\rho^{-1}\widehat L\rho\widehat L)-\Tr\widehat L^2.
\end{align}
In an eigenbasis
\(\rho=\operatorname{diag}(p_1,\ldots,p_d)\),
\begin{equation}
 K(\rho,L)
 =\sum_{i<j}\frac{(p_i-p_j)^2}{p_ip_j}|L_{ij}|^2\geq0,
 \label{eq:sm-Kpositive}
\end{equation}
with equality precisely when \([\rho,L]=0\).  Thus inefficiency turns
noncommutativity into a positive correction to \(dR/R\), thereby slowing the
decay of \(R\).  Hamiltonian rotations preserve the determinant.

\subsection{Full-rank preservation and expectation bound}

Fix a finite horizon \(t\), set
\(c_L=\|\widehat L\|_\infty\), and define
\[
 \tau_n=\inf\{u\geq0:\lambda_{\min}(\rho_u)\leq n^{-1}\}.
\]
Here and below \(n\) ranges over integers satisfying
\(n>\lambda_{\min}(\rho_0)^{-1}\).
Writing \(\widehat m_u=\Tr(\rho_u\widehat L_u)\), the drift in
Eq.~\eqref{eq:sm-logdet} is
\((1-\eta)K_u-2\eta dV_L-2\eta d\widehat m_u^2\).  Dropping \(K_u\geq0\)
gives, for \(v\leq t\),
\begin{equation}
 \ln\det\rho_{v\wedge\tau_n}
 \geq\ln\det\rho_0-C_Lt-\sup_{u\leq t}|\mathcal N_u|,
 \qquad
 C_L=2\eta d(V_L+c_L^2),
 \label{eq:sm-logdet-common-lower}
\end{equation}
where
\[
 \mathcal N_v=-2d\sqrt\eta\int_0^v\widehat m_u\,dW_u,
 \qquad
 \langle\mathcal N\rangle_v\leq4d^2\eta c_L^2v.
\]
The right side of Eq.~\eqref{eq:sm-logdet-common-lower} is finite almost
surely and independent of \(n\).  If rank were lost by time \(t\), continuity
would imply \(\tau_n\leq t\) for all large \(n\), together with
\(\ln\det\rho_{\tau_n}\leq-\ln n\).  This contradicts the finite common lower
bound.  A full-rank state
therefore remains full rank at every finite time.

Now set
\[
 Z_u^{(R)}=e^{2\eta V_Lu}R_u,\qquad
 \sigma_u=2\sqrt\eta(\mu-\Tr\rho_uL_u).
\]
Equation~\eqref{eq:sm-detroot-sde} becomes
\begin{equation}
 dZ_u^{(R)}=\frac{1-\eta}{d}Z_u^{(R)}K_u\,du
 +Z_u^{(R)}\sigma_u\,dW_u.
 \label{eq:sm-ZR-submartingale}
\end{equation}
On the fixed horizon, \(Z_u^{(R)}\leq e^{2\eta V_Lt}/d\) and
\(|\sigma_u|\leq2\sqrt\eta c_L\).  Full-rank preservation and eigenvalue
continuity also give a pathwise positive lower eigenvalue bound, so
\(\int_0^tZ_u^{(R)}K_u\,du<\infty\).  Stopping at
\[
 \zeta_k=\inf\left\{v:
 \int_0^vZ_u^{(R)}K_u\,du\geq k\right\}
\]
and taking expectations in Eq.~\eqref{eq:sm-ZR-submartingale} gives
\[
 \E Z_{t\wedge\zeta_k}^{(R)}
 =Z_0^{(R)}+\frac{1-\eta}{d}
 \E\int_0^{t\wedge\zeta_k}Z_u^{(R)}K_u\,du\geq Z_0^{(R)}.
\]
Dominated convergence as \(k\to\infty\) proves
Eq.~\eqref{eq:sm-detroot-bound}.  It also gives the equality statements,
because the nonnegative drift vanishes identically at \(\eta=1\), and at
\(\eta<1\) vanishes in expectation exactly when \(K=0\) almost everywhere.

\subsection{From determinant to impurity}

Let \(\gamma_d=(d-1)/d\), let \(p=\max_i p_i\), and put
\(\varepsilon=1-p\).  Since \(p\geq1/d\),
\[
 S=2\sum_{i<j}p_ip_j\geq2p\varepsilon\geq\frac{2\varepsilon}{d},
 \qquad
 \det\rho\leq
 \left(\frac{\varepsilon}{d-1}\right)^{d-1}.
\]
Consequently,
\begin{equation}
 S^{\gamma_d}\geq c_dR,\qquad
 c_d=\left[\frac{2(d-1)}d\right]^{\gamma_d}.
 \label{eq:sm-SR}
\end{equation}
For \(0<\theta\leq\gamma_d\), bounded-variable monotonicity and
Eq.~\eqref{eq:sm-detroot-bound} give
\begin{equation}
 \E[S_t^\theta]\geq c_dR_0e^{-2\eta V_Lt}.
 \label{eq:sm-detmoment-low}
\end{equation}
For \(\theta>\gamma_d\), put \(q=\theta/\gamma_d>1\), raise
Eq.~\eqref{eq:sm-SR} to the \(q\)th power, and apply Jensen's inequality to
obtain
\begin{equation}
 \E[S_t^\theta]\geq
 (c_dR_0)^q e^{-2q\eta V_Lt}.
 \label{eq:sm-detmoment-high}
\end{equation}
These bounds hold with the same prefactors for every
\(\pi\in\Pi_t^{(d)}\).  Taking the policy infimum therefore gives
\begin{equation}
 \Lambda_\star^{(d)}(\theta,\eta)
 \leq2\eta V_L\max\left\{1,\frac{\theta}{\gamma_d}\right\}.
 \label{eq:sm-detceiling}
\end{equation}

For \(V_L>0\), Popoviciu's inequality gives \(V_L\leq\ell^2\).  Its
range--variance equality requires the empirical spectral measure to place
equal weight at the two endpoints.  Thus equality requires an
even-dimensional balanced dichotomic spectrum.  All other spectra satisfy
\(V_L<\ell^2\).
For \(0<\eta<1\), comparison of Eq.~\eqref{eq:sm-detceiling} with the
attainable uniform ceiling in Eq.~\eqref{eq:sm-alld-piecewise} gives, whenever
\(V_L<\ell^2\), the strict boundary--interior separation interval
\begin{equation}
 \frac{1-\sqrt{1-V_L/\ell^2}}{2}
 <\theta<
 \gamma_d\frac{\ell^2}{V_L}.
 \label{eq:sm-detgapinterval}
\end{equation}
The two ceilings are strictly separated throughout this interval.

For the canonical qutrit
\(\operatorname{spec}L=(-\ell,0,\ell)\),
\(V_L=2\ell^2/3\), and
\begin{equation}
 \Lambda_\star^{(3)}(\theta,\eta)\leq
 \begin{cases}
 4\eta\ell^2/3,&0<\theta\leq2/3,\\[2pt]
 2\theta\eta\ell^2,&\theta>2/3.
 \end{cases}
 \label{eq:sm-qutritgap}
\end{equation}
At \(0<\eta<1\), the full-rank determinant ceiling lies strictly below the
attainable rank-two ceiling throughout
\[
 \frac{1-1/\sqrt3}{2}<\theta<1.
\]
For the sparse family
\(L_d=\operatorname{diag}(-\ell,0^{d-2},+\ell)\),
\(V_L=2\ell^2/d\), and the corresponding interval is
\[
 \frac{1-\sqrt{1-2/d}}2<\theta<\frac{d-1}{2}.
\]
These intervals quantify the boundary--interior separation for the canonical
qutrit and the sparse \(d\)-level family.

\subsection{Full qutrit spectral family}

After centering by a scalar shift and fixing the half-spread, write
\[
 L_\zeta=\ell\,\operatorname{diag}(-1,\zeta,1),\qquad
 -1\leq\zeta\leq1,\qquad
 v_\zeta=\frac{V_{L_\zeta}}{\ell^2}
 =\frac{2(3+\zeta^2)}9.
\]
For \(0<\eta<1\), exact comparison of the determinant and uniform ceilings
gives
\begin{equation}
 \theta_-(\zeta)<\theta<\theta_+(\zeta),\qquad
 \theta_-=\frac{1-\sqrt{1-v_\zeta}}2,\qquad
 \theta_+=\frac{2}{3v_\zeta}.
 \label{eq:sm-qutrit-family-band}
\end{equation}
The band ranges from
\(((1-1/\sqrt3)/2,1)\) at \(\zeta=0\) to
\((1/3,3/4)\) at \(|\zeta|=1\), with positive width across the full family.

\begin{figure}[t]
\centering
\includegraphics[width=0.92\textwidth]{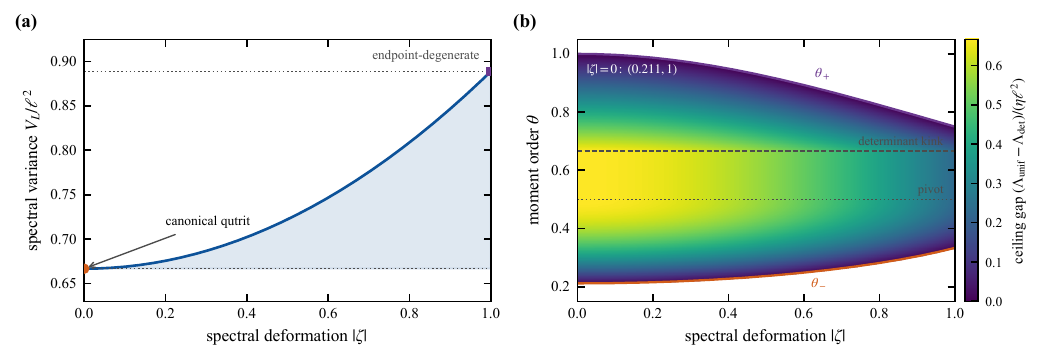}
\caption{Full-rank qutrit separation at \(0<\eta<1\) across
\(L_\zeta=\ell\,\operatorname{diag}(-1,\zeta,1)\).
(a) The normalized spectral variance
\(v_\zeta=2(3+\zeta^2)/9\).
(b) The positive part of the gap between the attainable state-space-uniform
ceiling and the determinant ceiling, in units of \(\eta\ell^2\).  It is
positive precisely between \(\theta_-(\zeta)\) and
\(\theta_+(\zeta)\).  Its positive width gives the proven strict separation.}
\label{fig:sm-qutrit-family}
\end{figure}

\section{Secondary finite-time resolutions}
\label{sm:secondary}

The following two exact asymptotic layers sharpen the physical interpretation
of the fixed-endpoint laws used above.

\subsection{Critical moment-order layer}
\label{sm:critical}

The change of QND saddle at \(\theta=1/2\) has a uniform boundary-layer
resolution.  Fix a mixed aligned initial state, let
\(v=4\eta\ell^2t\to\infty\), and set
\[
 \theta=\frac12+\frac{u}{2\sqrt v},\qquad u=O(1).
\]
Equation~\eqref{eq:sm-heat} gives, uniformly for \(u\) in compact sets,
\begin{equation}
 \frac{e^{v/2}\E_{y_0}[S_t^\theta]}{\sqrt{S_0}}
 =\mathcal F(u)+o(1),\qquad
 \mathcal F(u)=e^{u^2/2}
 \operatorname{erfc}\left(\frac{u}{\sqrt2}\right).
 \label{eq:sm-critical-scaling}
\end{equation}
Before the limit, the left side is
\[
 2^{-u/(2\sqrt v)}
 \E\!\left[
 \cosh(y_0+\sqrt v\,Z)^{-u/\sqrt v}
 \right],
 \qquad Z\sim N(0,1).
\]
Since \(v^{-1/2}\ln\cosh(y_0+\sqrt v\,Z)\to|Z|\), the limit is
\(\E[e^{-u|Z|}]=\mathcal F(u)\).  The identities
\(\mathcal F(0)=1\),
\(\mathcal F(u)\sim\sqrt{2/\pi}/u\) as \(u\to+\infty\), and
\(\mathcal F(u)\sim2e^{u^2/2}\) as \(u\to-\infty\) match the exact half-order
result,
the integrable spectral edge, and the two moving Gaussian saddles,
respectively.

\begin{figure}[t]
\centering
\includegraphics[width=0.92\textwidth]{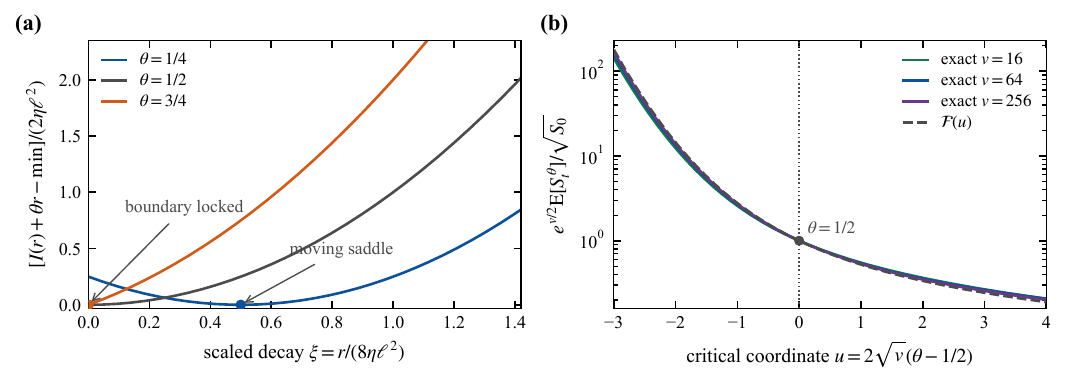}
\caption{Fixed-QND anatomy of the half-moment transition.
(a) The tilted large-deviation potential has minimizer
\(\xi_\theta=\max\{1-2\theta,0\}\).  The moving saddle reaches the physical
boundary at \(\theta=1/2\) and remains there.
(b) Finite-time moments at \(S_0=0.455\) collapse onto
\(\mathcal F(u)\) in Eq.~\eqref{eq:sm-critical-scaling}.  The curves use
quadrature of the exact QND kernel.}
\label{fig:sm-critical}
\end{figure}
\FloatBarrier

\subsection{Near-unit-efficiency crossover}
\label{sm:crossover}

The finite-time mean impurity remains continuous near the singular long-time
endpoint \(\eta=1\).  For an explicit comparison, put \(\delta=1-\eta\),
\(y_0=\operatorname{arctanh}\sqrt{1-2S_0}\), and
\(v=4\eta\ell^2t\).  Expanding the Gaussian in
Eq.~\eqref{eq:sm-heat} gives
\begin{equation}
 \E[S_t]_{\rm QND}
 =\frac{\sqrt{\pi S_0}}{2\sqrt v}e^{-v/2}
 \left[1-\frac{B_0}{v}+O(v^{-2})\right],
 \qquad
 B_0=\frac12\left(y_0^2+\frac{\pi^2}{4}\right).
 \label{eq:sm-qnd-tail-refined}
\end{equation}
The always-unbiased endpoint is Eq.~\eqref{eq:sm-ubbflow}, with
\(v/\eta=4\ell^2t\).
At the nonzero late crossing as \(\delta\downarrow0\), the transient in
Eq.~\eqref{eq:sm-ubbflow} is negligible relative to the floor
\(\delta/2\).  With
\[
 W_\delta=W_0\left(\frac{\pi S_0}{\delta^2}\right),
\]
where \(W_0\) is the principal Lambert function, the crossing obeys
\begin{align}
 v_\times
 &=W_\delta-\frac{2B_0}{1+W_\delta}
 +O(W_\delta^{-2}),
 \label{eq:sm-vcross}\\
 t_\times
 &=\frac{v_\times}{4\eta\ell^2}\nonumber\\
 &=\frac{2L_\delta-\ln(2L_\delta)+\ln(\pi S_0)
 +O(\ln L_\delta/L_\delta)}
 {4\eta\ell^2},
 \qquad L_\delta=\ln\frac1\delta.
 \label{eq:sm-across}
\end{align}
Thus the leading \(\ln(1/\delta)\) scale has a negative
\(\ln\ln(1/\delta)\) correction.

\begin{figure}[t]
\centering
\includegraphics[width=0.92\textwidth]{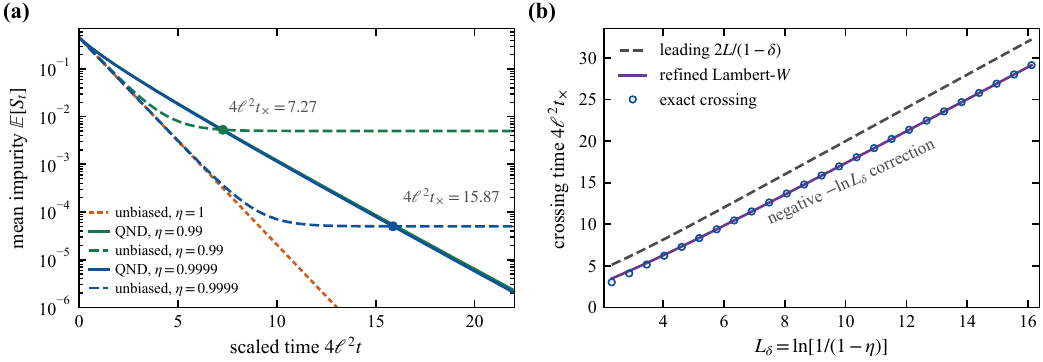}
\caption{Finite-time resolution of the singular efficiency limit, for
\(S_0=0.455\), plotted against the scaled time \(4\ell^2t\).
(a) Exact QND mean impurity (solid) and always-unbiased flow (dashed).
Circles mark their nonzero late crossings.
(b) Exact crossing times follow the refined Lambert-\(W\) expansion
including the negative \(\ln L_\delta\) correction.}
\label{fig:sm-crossover}
\end{figure}
\FloatBarrier

\section{Restoring laboratory-time units}
\label{sm:labunits}

Let \(T\) denote laboratory time and let a constant measurement strength
\(M_{\rm lab}>0\), with units of inverse time, multiply the fixed-spectrum
controlled channel.
After absorbing any separate finite-variation Hamiltonian into the relative
measurement orientation, the controlled-frame stochastic master equation is
\begin{equation}
\begin{aligned}
 d\rho={}&M_{\rm lab}\mathcal D[L_T^{\rm lab}]\rho\,dT\\
 &+\sqrt{\eta M_{\rm lab}}\,
 \mathcal H[L_T^{\rm lab}]\rho\,dW_T^{\rm lab},
 \qquad (dW_T^{\rm lab})^2=dT.
\end{aligned}
\label{eq:sm-lab-sme}
\end{equation}
With
\begin{equation}
 t=M_{\rm lab}T,\qquad
 dt=M_{\rm lab}dT,\qquad
 dW_t=\sqrt{M_{\rm lab}}\,dW_T^{\rm lab},\qquad
 L_t:=L_{t/M_{\rm lab}}^{\rm lab},
 \label{eq:sm-lab-map}
\end{equation}
this reduces to the unit-strength equation used throughout the Letter.
For each mapped admissible rule, a dimensionless exponent \(\Lambda\)
corresponds to the laboratory-time rate \(M_{\rm lab}\Lambda\).  The same
relation holds after optimizing over the time-horizon-dependent policy classes.

A prescribed, record-independent schedule \(M_{\rm lab}(T)\) admits the same
time change when it is strictly positive and locally integrable, with
\[
 \tau(T)=\int_0^T M_{\rm lab}(s)\,ds\longrightarrow\infty.
\]
One then takes
\[
 t=\tau(T),\qquad
 L_{\tau(T)}:=L_T^{\rm lab},\qquad
 W_{\tau(T)}=\int_0^T\sqrt{M_{\rm lab}(s)}\,dW_s^{\rm lab}.
\]
If the finite limit
\(\bar M=\lim_{T\to\infty}\tau(T)/T\in[0,\infty)\) exists, the corresponding
laboratory-time exponent equals \(\bar M\Lambda\).


\end{document}